\documentclass[conference,compsoc]{IEEEtran}
\ifCLASSOPTIONcompsoc
  \usepackage[nocompress]{cite}
\else
  \usepackage{cite}
\fi
\ifCLASSINFOpdf
\else
\fi
\usepackage{cite}
\usepackage{array}
\usepackage{multirow}
\usepackage{amsmath,amssymb,amsfonts}
\usepackage{algorithmic}
\usepackage{graphicx}
\usepackage{textcomp}
\usepackage{float}
\usepackage{longtable}
\usepackage{tabularray}
\usepackage{hyperref}
\usepackage{comment}
\usepackage{booktabs} 
\usepackage{subcaption}
\usepackage{makecell} 
\usepackage{braket}
\usepackage{xspace}
\usepackage{enumitem}
\usepackage{booktabs}
\usepackage{tabularx}  

\usepackage{placeins} 

\usepackage{array}      
\usepackage{tabularx}   
\usepackage{multirow}   
\usepackage{booktabs}   
\usepackage[table]{xcolor} 

\newcommand{\hlcell}[2][green!15]{\cellcolor{#1}#2}

\newcolumntype{L}[1]{>{\raggedright\arraybackslash}p{#1}}   
\newcolumntype{Y}{>{\raggedright\arraybackslash}X} 

\newenvironment{smitemize}{
  \begin{itemize}[topsep=1pt, partopsep=0pt, itemsep=1pt, parsep=0pt, leftmargin=10pt, itemindent=1pt]
}{\end{itemize}}

\newcommand{\paragraphB}[1]{\vskip 4pt \noindent \textit{\textbf{#1}.}\xspace}

\newcommand{\paragraphS}[1]{%
  \vspace{0.5ex}%
  \noindent{\fontsize{11}{13}\selectfont\bfseries #1.} %
}

\usepackage[color=green]{todonotes}

\usepackage{tikz}

\newcommand*\circled[1]{%
  \tikz[baseline=(char.base)]{
    \node[shape=circle, fill=black, inner sep=1pt] (char) {\scriptsize\textcolor{white}{#1}};}}

\usepackage[most]{tcolorbox}
\tcbset{
    colback=gray!10,       
    colframe=black!70,     
    boxrule=0.4pt,         
    arc=2pt,               
    left=6pt, right=6pt, top=4pt, bottom=4pt,
    fonttitle=\bfseries,
    coltitle=black
}

\definecolor{hl}{HTML}{FFF2CC} 
\usepackage{bbding}
\begin{document}
%
\title{SoK: Critical Evaluation of Quantum Machine Learning for Adversarial Robustness}



\renewcommand{\thefootnote}{\Envelope} 

\author{
\IEEEauthorblockN{
Saeefa Rubaiyet Nowmi\IEEEauthorrefmark{1}, 
Jesus Lopez\IEEEauthorrefmark{1},
Md Mahmudul Alam Imon\IEEEauthorrefmark{2},\\
Shahrooz Pouryousef\IEEEauthorrefmark{3},
Mohammad Saidur Rahman\IEEEauthorrefmark{1}
}

\IEEEauthorblockA{
\IEEEauthorrefmark{1}University of Texas at El Paso,
\IEEEauthorrefmark{2}University of Dhaka,
\IEEEauthorrefmark{3}University of Massachusetts Amherst
\\
\{srnowmi, jlopez126\}@miners.utep.edu,
mahmudulalam.imon@gmail.com,\\
shahrooz@cs.umass.edu,
msrahman3@utep.edu
}
}

\maketitle

\begin{abstract}
Quantum Machine Learning (QML) integrates quantum computational principles into learning algorithms, offering the potential for improved representational capacity and computational efficiency. Nevertheless, the security and robustness of QML systems remain largely underexplored, particularly under adversarial conditions. We present the first comprehensive systematization of adversarial robustness in QML, integrating conceptual organization with empirical evaluation across black-, gray-, and white-box threat models. We implement five representative attacks across all three threat models: a label-flipping data poisoning attack under black-box; an encoder-level indiscriminate data poisoning attack and
a proxy-model-based clean-label backdoor attack under gray-box; and a circuit-level backdoor attack (QTrojan) and gradient-based evasion attacks (FGSM and PGD) under white-box. We evaluate the attacks using a Quantum Multilayer Perceptron (QMLP) trained on MNIST and AZ-Class across circuit depths of 2, 5, 10 and 50 layers and two encoding schemes (angle and amplitude).

Our extensive evaluations reveal a fundamental accuracy–robustness trade-off. In particular, amplitude encoding yields the highest clean accuracy (92.6\% on MNIST, 67\% on AZ-Class); however, it collapses under adversarial perturbations and depolarizing noise, while shallow angle-encoded models remain substantially more stable. In addition, QUID is highly effective under noiseless conditions but is weakened by noise, whereas the proxy-model backdoor persists unless the circuit itself is overwhelmed, highlighting that noise is an asymmetric and unreliable passive defense. Furthermore, the circuit-level backdoor fails in the multi-class setting, indicating a scalability constraint. Finally, QMLP models are more robust than Classical Multi-Layer Perceptron (CMLP) models under label-flipping attacks but are substantially more vulnerable to gradient-based evasion, motivating the need for quantum-specific defenses. We conclude by proposing a threat-aware, noise-resilient design framework for secure and robust QML deployment.
\end{abstract}


\paragraphB{Reproducibility}
Code is available at \url{https://iqsec-lab.github.io/SoK-QML/}. 


\IEEEpeerreviewmaketitle

\section{Introduction}

\begin{figure}[!t]
\centering
    \includegraphics[width=\linewidth]
    {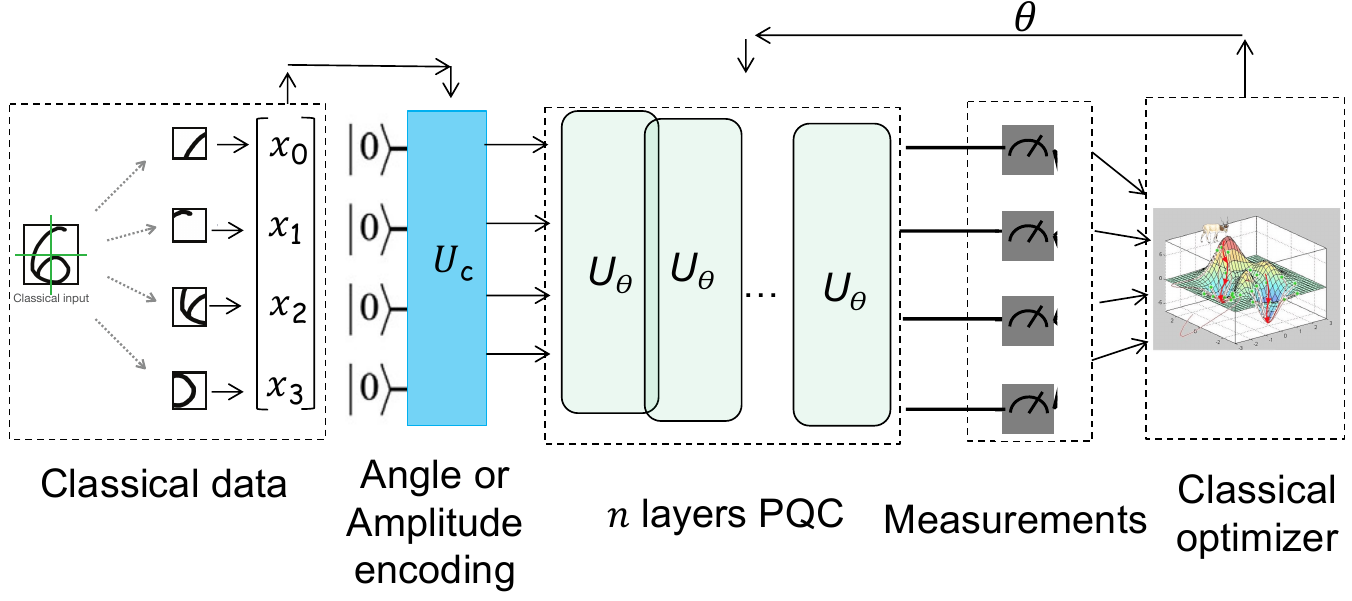}
    \caption{Architecture of Parameterized Quantum Circuit (PQC) as a QML model. 
    The circuit comprises three key components: (1) data encoding layers that map classical inputs into quantum states, (2) parameterized quantum gates that define the model’s behavior and are iteratively optimized, and (3) measurement operations that extract classical information for loss evaluation. 
  }\label{fig:VQC}
  \vspace{-0.3cm}
\end{figure}

QML leverages fundamental quantum phenomena such as superposition, entanglement, and interference to explore potential for {\em quantum advantage} over Classical Machine Learning (CML)\cite{ren2022experimental,huang2022quantum}. QML techniques have been investigated across diverse domains including combinatorial optimization~\cite{ye2023towards}, cybersecurity~\cite{lopez2025towards}, drug discovery \cite {Barta2021Drug} and so on. As QML transitions from theoretical exploration to practical implementation  \cite{ren2022experimental,huang2022quantum}, the absence of a systematic understanding of adversarial robustness poses a critical barrier to their trustworthy deployment. 

The emergence of cloud-based quantum computing services such as IBM Quantum\cite{ibmq}, Amazon Braket\cite{amazonbraket}, and Microsoft Azure Quantum\cite{azurequantum} has accelerated the accessibility and scalability of QML. However, it has also broadened the attack surface~\cite{kundu2025stiq,kundu2024}. 
Multi-tenant infrastructures, remote execution APIs, and shared quantum backend introduce novel vulnerabilities including crosstalk~\cite{choudhury25crosstalk}, data leakage~\cite{lu2025quantum}, and side-channel exposure~\cite{lu2025quantum}, undermining the assurance of confidentiality and model integrity. Moreover, the hybrid architecture of QML as illustrated in Figure \ref{fig:VQC}, which couples quantum and classical components, introduces complex dependencies across the software-hardware stack~\cite{Lubinski2022Advancinghybrid}. 
As such, QML systems inherit CML vulnerabilities and exhibit poorly understood quantum-specific threats.

Recent studies have showed that QML models are vulnerable to both {\em training}  and {\em inference} time attacks including evasion~\cite{gong2022universal,west2023benchmarking}, poisoning~\cite{Kundu-poisoning-2025}, backdoor~\cite{zhao2025backdoor}. In addition to the algorithmic attacks, hardware-level vulnerabilities have also been identified in Noisy Intermediate-Scale Quantum (NISQ) systems, such as state-leakage~\cite{Jakub2024StateLeakage,tan2025qubithammerattacksqubitflipping}, crosstalk~\cite{choudhury25crosstalk, zhao2023mitigation}, power and timing side channel leakage\cite{lu2025quantum,xu2023securing}, and pulse level manipulations~\cite{xu2024quantum,xu2025security}. These findings expose the fragility of QML's security landscape and highlight the urgent need for a {\em systematic understanding of adversarial robustness in quantum learning systems}.


To address this gap, we propose a Systematization of Knowledge (SoK) that consolidates the research on QML adversarial robustness, develops an empirical framework for evaluating real-world vulnerabilities, and outlines a road-map towards secure and resilient QML. Towards that end, this work is driven by three fundamental research questions for systematization and empirical validation:

\begin{enumerate}[label=\protect\circled{RQ\arabic*}, leftmargin=*]

    \item What are the dominant threat vectors in hybrid quantum-classical architectures, and how can they be systematically classified to capture both classical and quantum-specific adversarial capabilities?
    
    \item How do {\em circuit depth} and {\em data encoding} influence the performance and adversarial robustness of {\em Parameterized Quantum Circuits} (PQCs) within the constraints of NISQ-era hardware?
    
    \item To what extent are the current QML models and defense mechanisms resilient against respective attacks spanning black-box, gray-box, and white-box threat models compared to CML models?
\end{enumerate}

Addressing these questions is critical for bridging the gap between theoretical advancement in QML and its secure, practical deployment. To this end, we structure our study in four stages. First, to explore \protect\circled{RQ1}, we construct a hybrid quantum-classical pipeline representative of model Quantum-as-a-Service (QaaS) platforms (e.g., IBM Quantum, Amazon Braket, and Microsoft Azure Quantum). These multi-tenant cloud environments inherently expand the attack surface by combining classical and quantum components. As such, we develop a {\em taxonomy of QML threat models} categorizing adversarial capabilities into black-box, gray-box, and while-box levels, to enable systematic reasoning about vulnerabilities across the hybrid stack (\textbf{Section~\ref{sec:taxonomy}}).

Secondly, to investigate \protect\circled{RQ2}, we analyze Quantum Multilayer Perceptron (QMLP) with two canonical data encoding schemes- angle and amplitude encoding and varying circuit depths (2, 5, 10, and 50 layer PQCs), to realize how architectural parameters shape accuracy, efficiency, and adversarial robustness under NISQ constraints (\textbf{Section~\ref{sec:QMLPBaseline}}).
To answer \protect\circled{RQ3}, we empirically implement at least one representative attack from each threat model category to evaluate the resilience of QML models and the effectiveness of existing defenses (\textbf{Section~\ref{sec:label-flipping},~\ref{sec:quid-eval}, and~\ref{sec:PGD-FGSM}}). 

These attack-defense implementations are chosen for their prevalence and demonstrated impact on prior QML security research~\cite{lu2020quantum,west2023benchmarking,gong2022universal,Kundu-poisoning-2025}. From the black-box threat model, we examine the {\em data poisoning (label flipping)}\cite{Bhatia2024QFL-LF} attack and apply \emph{label smoothing} \cite{Szegedy2015RethinkingTI} as its defense mechanism. For the gray-box threat model, we implement the \emph{Quantum Indiscriminate Data Poisoning (QUID)} attack~\cite{Kundu-poisoning-2025} and evaluate the \emph{Q-Detection} defense~\cite{he2025q}. We also evaluate a backdoor attack under gray-box threat model~\cite{huang2023backdoor}. 
Finally, under the white-box threat model, we examine gradient-based adversarial evasion attacks using \emph{Projected Gradient Descent (PGD)}~\cite{madry2018towards} and the \emph{Fast Gradient Sign Method (FGSM)}~\cite{goodfellow2014explaining}, as well as a circuit trojan based, Qtrojan~\cite{chu2023qtrojan}. 
We do not replicate the attacks and defenses exactly as presented in the literature; instead, we adapt them to our experimental framework to ensure consistency and comparability across settings.

Finally, we distill our empirical findings into {\em practical design principles} for developing secure, noise-aware, and resilient QML pipelines tailored to NISQ-era constraints, bridging the gap between theoretical potential and real-world security assurance. Given the growing importance of secure quantum learning systems, this work presents both a systematization and an empirical investigation of adversarial robustness in QML. 

Our main contributions are summarized as follows:

\begin{itemize}
    \item We present a comprehensive systematization of adversarial attack models in QML, categorized by the adversary’s level of access.
    
    \item We empirically analyze how data encoding schemes 
    (angle vs. amplitude) and circuit depth (2, 5, 10, and 50 layers) influence QML performance and robustness under NISQ constraints. 
    We find that under noiseless conditions, amplitude-encoded deep QMLP models outperform angle-encoded QMLP, whereas under noisy conditions, shallow angle-encoded QMLPs exhibit superior robustness.

    \item We perform a comparative study with classical machine learning, such as CMLP (Classical Multilayer Perceptron), to highlight distinctive vulnerability patterns and motivate the need for quantum-specific defenses. 
    
    \item We implement representative attacks and corresponding defenses across black-box, gray-box, and white-box threat models to evaluate the robustness of QML models, compare against to CML models, and assess the effectiveness of existing defense mechanisms. We find that QMLP models are more robust against classical label-flipping attacks but substantially more vulnerable to gradient-based perturbation attacks.

    \item We propose a structured design pipeline for developing secure and robust QML models integrating threat modeling, encoding strategy selection, and robustness evaluation under quantum noise.
\end{itemize}

To ground our systematization, we curate works from leading security venues--IEEE S\&P, ACM CCS, USENIX Security, and NDSS, published between 2020 and 2025, complemented by research from major quantum such as Nature, Science, Nature Reviews Physics, npj Quantum Information, Physical Review, and machine learning venues such as NeurIPS, IJCAI, ICML.



\paragraphS{QML Robustness Study as SoK} 
While quantum computing and QML remain nascent, we argue that this SoK is both timely and necessary for several reasons. 
First, adversarial attacks on QML are no longer merely theoretical. Prior works have demonstrated that targeted attacks on QML models are practically feasible even without physical hardware ownership~\cite{choudhury25crosstalk,Fu2024QuantumLeak}. Second, QML inherits a substantial attack surface from CML~\cite{gong2022universal, west2023benchmarking}, yet their behavior in the quantum setting is underexplored, requiring systematic study to determine where classical insights transfer and where quantum-specific effects demand new defenses. Third, robustness research in QML is fragmented across the physics and computer science communities. Physics-oriented work emphasizes theoretical guarantees under ideal assumptions \cite{weber2021optimal,Angrisani2023DiffrentialPrivacy,barooti2021provable,gong2024enhancing}, while computer science research focuses on empirical evaluation under realistic NISQ constraints~\cite{west2023benchmarking,choudhury25crosstalk}. A SoK, is therefore, precisely needed to bridge these perspectives into a unified, coherent, and actionable framework for secure QML deployment.

\section{Preliminaries}
\label{sec:Preliminaries}
\vspace{-0.3cm}
In this section, we first review the fundamentals of quantum computation, including qubits, quantum gates, and their mathematical representations, and then summarize key QML concepts, including hybrid quantum-classical models and variational circuits, and the noise and limitations in current NISQ devices.

\paragraphS{Qubits and Gates}
Quantum computing operates on \emph{qubits}, the quantum counterparts of classical bits. Unlike classical bits that are either \(0\) or \(1\), a qubit can exist in superposition of states:

\begin{equation}
    \ket{\psi} = \alpha \ket{0} + \beta \ket{1},
\end{equation}

where \( \alpha, \beta \in \mathbb{C} \) and \( |\alpha|^2 + |\beta|^2 = 1 \). Measurement collapses the state to \( \ket{0} \) or \( \ket{1} \) with corresponding probabilities \( |\alpha|^2 \) and \( |\beta|^2 \). Quantum operations are performed using \emph{unitary gates}, which preserve normalization. For an \(n\)-qubit system, a gate \( U \in \mathbb{U}(2^n) \) satisfies:

\begin{equation}
    U^\dagger U = U U^\dagger = I,
\end{equation}

where \( U^\dagger \) is the Hermitian adjoint and \( I \) is the identity. Common single-qubit gates include \( X, Y, Z, H, R_X(\theta), R_Y(\theta), R_Z(\theta) \), while multi-qubit gates such as CNOT and CRX enable entanglement, an essential feature for quantum computation~\cite{nielsen2010quantum}.

\paragraphS{Quantum Machine Learning (QML)}
QML integrates quantum computing (QC) and machine learning (ML) to improve how algorithms learn from data~\cite{biamonte2017quantum}. A quantum state is represented as a vector of $\ket{\psi}$ in a $2^n$-dimensional space for $n$ qubits, where information can exist in multiple states at once, a property known as {\em superposition}. Quantum effects such as {\em entanglement} and {\em interference} allow computations that can explore many possibilities simultaneously~\cite{nielsen2010quantum}. This parallelism offers potential speedups for problems like optimization and pattern recognition that are central to ML~\cite{peters2021machine}. 

Most practical QML systems use a {\em hybrid quantum-classical} approach. As shown in Figure~\ref{fig:VQC}, the quantum processor executes {\em Parameterized Quantum Circuits} (PQCs) that prepares parameterized quantum circuits that prepare quantum states $\ket{\psi(\boldsymbol{\theta})} = U(\boldsymbol{\theta})\ket{0}^{\otimes n}$, where $\boldsymbol{\theta} = (\theta_1,\ldots,\theta_\ell)$ are trainable parameters and $U(\boldsymbol{\theta})$ is unitary~\cite{benedetti2019parameterized}. Measurement outcomes are sent to a classical optimizer which updates $\boldsymbol{\theta}$ to minimize a loss $L(\boldsymbol{\theta})$ via 

\begin{equation}
    \boldsymbol{\theta}^{(t+1)} = \boldsymbol{\theta}^{(t)} - \eta \nabla_{\boldsymbol{\theta}} L(\boldsymbol{\theta}^{(t)}),
\end{equation}


with learning rate $\eta$. The measurement-optimization process forms a {\em variational quantum circuit} (VQC) and is the foundation of algorithms such as variational quantum eigensolver (VQE)~\cite{kandala2017hardware}, quantum neural network~\cite{cerezo2021variational}, and quantum convolutional neural network (QCNN)~\cite{cong2019quantum}. 

A typical QML pipeline consists of three stages-data embedding, parameterized quantum gates, and measurement followed by classical postprocessing (See Figure~\ref{fig:VQC}).

\subsubsection*{Data Embedding}
Classical input representation must be mapped into quantum states before processing by a QML model. This step, called {\em data embedding} or {\em quantum feature mapping}, is performed using parameterized quantum gates such as rotation gates $R_X$ or $R_Y$~\cite{lloyd2020quantum}. The choice of embedding method directly affects performance, with trade-offs in qubit count, circuit depth, and noise. Two common approaches are \emph{angle encoding} and \emph{amplitude encoding}.


\begin{smitemize}
    \item \textbf{Angle Encoding:} In this scheme, each classical feature $x_i$ is mapped to a rotation angle of a single-qubit gate~\cite{dowling2024adversarial}. For an input vector \( \vec{x} \in \mathbb{R}^N \), rotation gates \( R_X(x_i) \) or \( R_Y(x_i) \) are applied to \( N \) qubits to prepare quantum state:
    \[
        R_Y(x_i)\ket{0} \;\text{or}\; R_X(x_i)\ket{0}.
    \]
    This method is simple and hardware-efficient but requires at least one qubit per input feature (\( N \leq n \)).

    \item \textbf{Amplitude Encoding:} In this method, the entire input vector is encoded into the amplitudes of a quantum state~\cite{dowling2024adversarial}:
    \begin{equation}
        \ket{\phi} = \sum_{i=0}^{N-1} x_i \ket{i},
    \end{equation}
    where \( \ket{i} \) are basis states of \( \log_2 N \) qubits. This method is qubit-efficient but harder to implement on NISQ hardware due to complex state preparation and noise sensitivity.
\end{smitemize}

\subsubsection*{Parameterized Quantum Gates}
After input embedding, the circuit applies trainable unitary operations called \emph{variational layers}~\cite{zomorodi2024optimal}. Each layer consists of parameterized single-qubit rotation gates $R_X(\theta)$, $R_Y(\theta)$, and $R_Z(\theta)$, combined with multi-qubit entangling gates such as \texttt{CRX} or \texttt{CNOT}. These gates form a PQC where parameters \( \boldsymbol{\theta} \) are initialized randomly and updated during training, enabling the model to learn transformations while maintaining unitarity.

\subsubsection*{Measurement and Postprocessing}
After variational layers, the quantum circuit is measured in computational basis to obtain classical outputs. Each measurement corresponds to expectation value of a Pauli-\( Z \) operator~\cite{peres2023pauli}.

\begin{equation}
    z_i = \bra{\psi(\boldsymbol{\theta})} Z_i \ket{\psi(\boldsymbol{\theta})},
\end{equation}

where $z_i \in [-1, 1]$ represents the outcome associated with qubit $i$. The resulting vector $\mathbf{z} = (z_1, \ldots, z_n)$  forms the quantum feature representations passed to classical post-processing layers such as fully connected or softmax classifiers. Gradients with respect to circuit parameters $\boldsymbol{\theta}$ are computed using the parameter-shift rule or finite difference methods, and classical optimizers (e.g., Adam, SGD) update the parameters in a hybrid training loop~\cite{kohda2022quantum,patel2025quantum}.



\paragraphS{Quantum Multilayer Perceptron (QMLP)}
\label{prelem:QMLP}
QMLP is a hybrid quantum-classical model for supervised learning that integrates PQC within a classical workflow~\cite{biamonte2017quantum}. Classical inputs \( \mathbf{x} \in \mathbb{R}^d \) are first encoded into quantum states using either \emph{angle encoding} or \emph{amplitude encoding}. To enhance representational robustness, same features can be re-encoded across layers through re-uploading, analogous to stacking layers in classical neural networks. Following embedding, the state passes through variational layers with trainable single-qubit rotations and \texttt{CRX} entangling gates. The quantum output is measured via Pauli-\( Z \) expectation values producing a real-valued vector \( \mathbf{z} \in \mathbb{R}^n \) that serves as input to classical fully connected layer. The model is trained end-to-end using gradient-based optimization in a hybrid quantum-classical loop.

\paragraphS{Quantum Noise and NISQ Constraints}
Current NISQ devices suffer from decoherence, limited gate fidelity, and readout noise, all of which degrade circuit performance~\cite{preskill2018quantum}. 
These imperfections are commonly modeled as \emph{quantum noise channels} acting on a quantum state $\rho$. 
A general noisy process is expressed as a completely positive trace-preserving (CPTP) map:
\begin{equation}
    \mathcal{E}(\rho) = \sum_k E_k \rho E_k^\dagger,
\end{equation}
where $\{E_k\}$ are Kraus operators satisfying $\sum_k E_k^\dagger E_k = I$. Typical single-qubit noise channels include:

\begin{smitemize}
    \item \textbf{Depolarizing noise:}
    \begin{equation}
        \mathcal{E}_p(\rho) = (1-p)\rho + \frac{p}{3}(X\rho X + Y\rho Y + Z\rho Z),
    \end{equation}
    which replaces the qubit with a maximally mixed state with probability $p$.

    \item \textbf{Amplitude damping:} models energy relaxation with Kraus operators
    \[
        E_0 = 
        \begin{bmatrix}
            1 & 0 \\
            0 & \sqrt{1-\gamma}
        \end{bmatrix},
        \quad
        E_1 = 
        \begin{bmatrix}
            0 & \sqrt{\gamma} \\
            0 & 0
        \end{bmatrix}.
    \]
\end{smitemize}

Noise accumulates with circuit depth, reducing fidelity and compounding the effects of adversarial perturbations. 


\section{Taxonomy of Attack Models}
\label{sec:taxonomy}

\begin{figure}[!t]
    \centering
    \includegraphics[width=\linewidth]{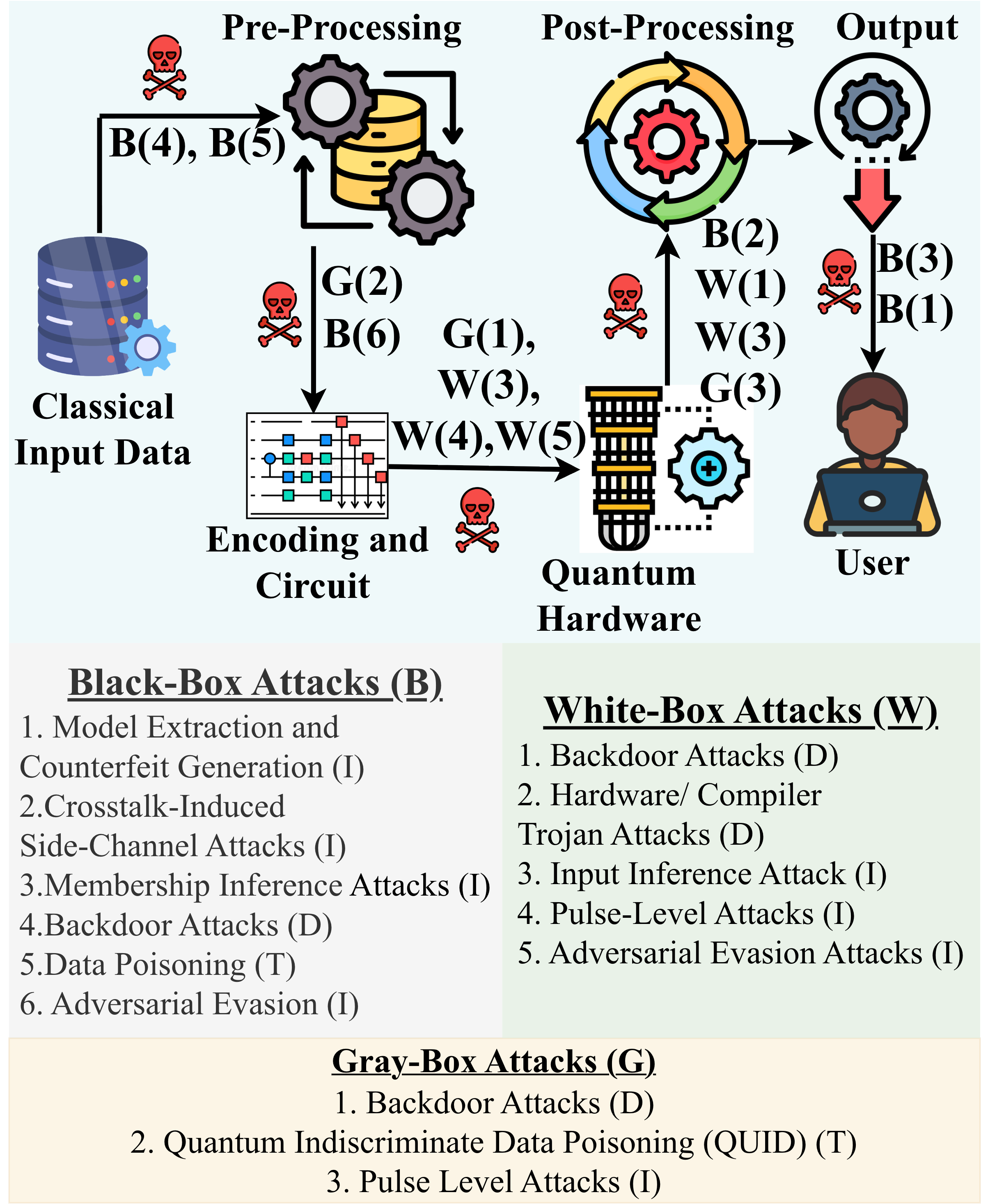}
    \caption{Schematic of adversarial attack surfaces in QML pipeline, illustrating potential threat vectors and their points of insertion. 
    Each attack is annotated as \textbf{(T)}: Training-time, \textbf{(I)}: Inference-time, \textbf{(D)}: Dual-phase (embedded during training, activated during inference), \textbf{(B)}: Black-Box, \textbf{(G)}: Gray-Box, \textbf{(W)}: White-Box.}
    \label{fig:Taxonomy}
    \vspace{-0.3cm}
\end{figure}

\begin{table*}[!t]
\centering
\footnotesize
\setlength{\tabcolsep}{2pt}
\caption{Systematization of adversarial attacks on QML classified by threat model (access level) and attack phase. T: Training-time attack, I: Inference-time attack. Attacks empirically evaluated in this study are highlighted in green.}
\label{tab:Attack_taxonomy}
\renewcommand{\arraystretch}{1.25}
\begin{tabularx}{\textwidth}{L{1.3cm}|L{3.05cm}|L{1.5cm}|L{0.5cm}|L{0.5cm}|L{2.2cm}|L{3.75cm}|L{3.5cm}}
\toprule
\textbf{Threat} \newline \textbf{Model}
& \textbf{Attack Category}
& \textbf{Related} \newline \textbf{Paper}
& \textbf{T} 
& \textbf{I}
& \textbf{Attacker} \newline \textbf{Profile}
& \textbf{Target Artifacts} 
& \textbf{Capabilities} \\
\midrule
\multirow{6}{*}{\parbox{1.3cm}{Black-box}}
& Model Extraction \& Counterfeit Generation
& \cite{kundu2024evaluating,Fu2025CopyQNN,Fu2024QuantumLeak}
& & \checkmark
& \multirow{6}{=}{External adversary, co-tenant on multi-tenant NISQ hardware, malicious data contributor}
& \multirow{6}{=}{Third-party Training Data, QML Outputs}
& \multirow{6}{=}{Model Stealing, Functionality replication, Poisoning the data} \\
\cline{2-5}
& Crosstalk-induced Side-Channel
& \cite{choudhury25crosstalk,xu2025security}
& & \checkmark & & & \\
\cline{2-5}
& Membership Inference
& \cite{watkins2023quantum}
& & \checkmark & & & \\
\cline{2-5}
& \hlcell{Black-Box Backdoor}
& \cite{zhao2025backdoor,Bhowmik2025QuPT}
& \checkmark & \checkmark & & & \\
\cline{2-5}
& \hlcell{Data Poisoning (Label-Flipping)}
& \cite{Bhatia2024QFL-LF,Szegedy2015RethinkingTI}
& \checkmark & & & & \\
\cline{2-5}
& Adversarial Evasion
& \cite{gong2022universal}
& & \checkmark & & & \\
\midrule
\multirow{3}{*}{\parbox{1.3cm}{Gray-box}}
& \hlcell{Gray-Box Backdoor} 
& \cite{chu2023qdoor, huang2023backdoor}
& \checkmark & \checkmark
& \multirow{3}{=}{Semi-privileged adversary, cloud provider}
& \multirow{3}{=}{Transpiled circuits, pre-execution artifacts, training data, pulse schedules, neighboring qubits}
& \multirow{3}{=}{Manipulate inputs, training data, or intermediate representations; Influence model behavior} \\ \cline{2-5}
& \hlcell{QUID}
& \cite{Kundu-poisoning-2025}
& \checkmark & & & & \\
\cline{2-5}
& Pulse-level Attacks
& \cite{xu2025security, Shubha2025Pulse}
& & \checkmark & & & \\
\midrule
\multirow{5}{*}{\parbox{1.3cm}{White-box}}
& \hlcell{{{White-Box Backdoor}}}
& \cite{chu2023qtrojan, guo2025backdoor}
& \checkmark & \checkmark
& \multirow{5}{=}{Insider with full access (transpiler/ infrastructure insider)}
& \multirow{5}{=}{Transpiled gates, PQC, pulse schedule; parameters and gradients}
& \multirow{5}{=}{Analyze and RE circuits, inject Trojan into hardware, Poison data in training time} \\
\cline{2-5}
& Hardware/ Compiler Trojan
& \cite{roy2024hardwaretrojans}
& \checkmark & \checkmark & & & \\
\cline{2-5}
& Input Inference
& \cite{Heredge2025}
& & \checkmark & & & \\
\cline{2-5}
& Reverse Engineering
& \cite{Ghosh2024Imitation,ghosh2025ai}
& & \checkmark & & & \\
\cline{2-5}
& \hlcell{{{Adversarial Evasion (FGSM/PGD)}}}
& \cite{west2023towards,west2023benchmarking}
& & \checkmark & & & \\
\bottomrule
\end{tabularx}
\end{table*}

In this section, we address \protect\circled{RQ1} by presenting a \emph{taxonomy} of adversarial threats in QML based on the adversary's level of access-black-box, gray-box, and white-box, and the quantum-specific mechanisms exploited during the attack. 
While prior works have explored individual attack scenarios \cite{lu2020quantum,Kundu-poisoning-2025,choudhury25crosstalk}, the absence of a unified framework has limited our understanding of how vulnerabilities propagate across different stages of the QML pipeline. 
We address this gap through our taxonomy, which highlights how each threat model maps to distinct attack categories such as \emph{encoding manipulation}\cite{Kundu-poisoning-2025}, \emph{circuit tampering}, \cite{west2023benchmarking}and \emph{hardware level side-channel attacks}\cite{choudhury25crosstalk}. 
This systematic classification also enables consistent comparison across existing literature and clarifies how classical attack paradigms extend into the quantum domain. 
Figure~\ref{fig:Taxonomy} illustrates the attack vectors within the QML workflow, and Table~\ref{tab:Attack_taxonomy} summarizes representative attacks under each threat model. 

To operationalize this framework, we decompose the adversarial landscape along three key dimensions--{\em the adversary’s level of access}, {\em the nature of the exploited mechanism} and {\em the phase of attack deployment (i.e., training or inference)}.
The first dimension, {\em access level}, captures the extent of the adversary’s visibility and control over the QML workflow, defining three canonical threat models: {\em black-box} (i.e., query-only and third-party data access), {\em gray-box} (i.e., partial visibility into intermediate artifacts such as encoded states or transpiled circuits), and {\em white-box} (i.e., full access to parameters, architecture, and hardware).
The second dimension distinguishes whether the attack leverages {\em classical-style manipulations} (e.g., data poisoning or gradient-based evasion) or {\em quantum-specific mechanisms} such as encoding interference, entanglement manipulation, or pulse-level perturbations. 
The third dimension, {\em phase of attack deployment}, distinguishes whether the attack targets the {\em training stage} (e.g., label-flipping, QUID\cite{Kundu-poisoning-2025}, and backdoor injection via quantum universal perturbations \cite{Zhao_2025_QUAP}) or the {\em inference stage} (e.g., FGSM, PGD \cite{lu2020quantum,west2023benchmarking}, and circuit-level backdoor attacks such as QTrojan \cite{chu2023qtrojan}), as each phase exposes different system artifacts to compromise and demands distinct defense strategies.

This three-dimensional framework provides a unified lens for comparing classical and quantum vulnerabilities, illustrating how QML both inherits and extends traditional ML threats while introducing fundamentally new attack surfaces arising from quantum computation.

\subsection{Black-Box Attack}
\label{subsec:Black-box}

In the {\em black-box} setting, the adversary lacks visibility into the QML model’s architecture, parameters, and hardware. 
Access is limited to model queries, as is typical of cloud-hosted QML-as-a-Service (QMLaaS) or indirect influence on third-party data sources. 
Despite this {\em lowest level of access}, attackers can still exploit both {\em classical-style manipulations} \cite{gong2022universal} and {\em quantum-specific mechanisms} \cite{zhao2025backdoor} through system interfaces or shared resources~\cite{Kundu-poisoning-2025,choudhury25crosstalk}.


\paragraphB{Model Extraction and Counterfeit Generation}
A query-only adversary can approximate a target QML model by generating input-output pairs and training a local surrogate using these observations~\cite{kundu2024evaluating,Fu2025CopyQNN,Fu2024QuantumLeak}. This classical-style manipulation enables functional replication and intellectual-property theft, allowing the attacker to reproduce decision boundaries or redeploy counterfeit models. Such attacks compromise confidentiality and expose QMLaaS providers to significant intellectual-property and usage risks.

\paragraphB{Crosstalk-induced side-channel attacks}
On multi-tenant NISQ devices, physical interference between qubits allows a co-resident attacker to observe correlated disturbances in {\em sensing} qubits and infer circuit activity or timing~\cite{xu2025security,choudhury25crosstalk}. This is a quantum-specific mechanism that reveals structural or operational metadata about the victim’s workload without direct access to its circuit or data. This vector principally undermines Confidentiality (leakage of circuit design and operational metadata) and can have secondary consequences for Integrity if reconstructed designs are misappropriated.

\paragraphB{Membership Inference Attacks}
In this black-box privacy threat, an adversary determines whether data sample was used during model training by comparing model’s confidence or output distribution across queries~\cite{watkins2023quantum}. Such inference exploits overfitting and distributional biases to expose membership information, violating confidentiality even when models are deployed as restricted-access services.


\paragraphB{Black-Box Backdoor Attacks (BBBAs)}
BBBAs assume little to no knowledge of the target model and typically exploit transferability or vulnerabilities in the quantum computing stack.
For example, Quantum Properties Trojans (QuPT)~\cite{Bhowmik2025QuPT}
introduces malicious functionality at the circuit or software level without
access to training data or model parameters. 
This represents a shift toward supply-chain-style attacks, where vulnerabilities arise from the broader quantum software stack.
Recent work also uses quantum universal adversarial perturbations (QUAP)~\cite{zhao2025backdoor} to realize BBBAs. QUAP generates transferable perturbations that can serve
as trigger patterns in
poisoned training data, enabling
high attack success rates across different QNN architectures
even under strict black-box assumptions. 

\paragraphB{Data Poisoning (Label-Flipping)}
In a purely classical poisoning variant, the adversary contaminates third-party datasets by flipping labels for a subset of samples while leaving features unchanged~\cite{Bhatia2024QFL-LF}. This training-time corruption degrades model accuracy and reliability, leading to global misclassification errors. We evaluate this in Section~\ref{sec:label-flipping}, analyzing how encoding schemes and circuit depth influence robustness under noiseless and noisy conditions.

\paragraphB{Adversarial Evasion Attack (AEA)} 
AEA exploits perturbation transferability to attack models without direct access. A key example is universal adversarial perturbations for quantum classifiers, where a single perturbation generalizes across inputs and models. As shown in~\cite{gong2022universal}, these perturbations can deceive multiple quantum classifiers simultaneously, revealing persistent adversarial vulnerabilities. This behavior suggests shared decision boundaries or feature representations, enabling practical black-box evasion attacks.

\subsection{Gray-Box Attack}
\label{subsec:Gray-box}
In the gray-box setting, adversaries have partial visibility into the QML system, including access to intermediate representations such as encoded quantum states, transpiled circuits, and compiler configuration files, but not the full model parameters or hardware. 
This setting commonly arises in shared or outsourced environments such as compromised QaaS platforms or insider threats, where attackers can observe or slightly modify parts of the pipeline \cite{Kundu-poisoning-2025,wang2023qumos}. 
Within this threat model, adversaries can combine classical control (e.g., data manipulation or configuration changes) with quantum- specific mechanisms (e.g., circuit injection or state perturbation) to compromise integrity or extract sensitive information.

\paragraphB{Backdoor Attacks} Gray-box backdoor attacks assume partial knowledge of the target model, often leveraging surrogate models or limited pipeline access. 
QDoor exemplifies this setting by embedding a backdoor during training through the quantum compilation process, where approximate synthesis as a conditional trigger, without requiring deployment-time access~\cite{chu2023qdoor}. 
Similarly, Huang et al.~\cite{huang2023backdoor} propose a data poisoning backdoor that uses a proxy model to generate transferable triggers without full knowledge of the victim QNN, showing that backdoor attacks remain effective under limited knowledge.


\paragraphB{Pulse-level Attacks} 
By exploiting low-level control interfaces to manipulate pulse parameters (e.g., timing, amplitude, and waveform), attackers can influence neighboring qubits through hardware coupling effects with pulse-level attacks.
As shown in \cite{xu2025security,Shubha2025Pulse}, such attacks do not require access to the victim’s circuit; instead leveraging crosstalk to induce targeted errors. Operating below the gate-level abstraction, they exploit physical-layer dynamics for stealthy and indirect manipulation, highlighting a distinct gray-box threat surface arising from cross-qubit interactions and abstraction-layer inconsistencies.

\paragraphB{Quantum Indiscriminate Data Poisoning Attack (QUID)}
The QUID attack~\cite{Kundu-poisoning-2025} illustrates how partial access to the quantum encoder suffices to corrupt learning. It exploits {\em Encoder State Similarity (ESS)}- the geometric closeness of quantum states in Hilbert space, to perturb labels and induce class confusion during training~\cite{Watrous_2018,Nielsen_Chuang_2010,Kundu-poisoning-2025}.
Unlike gradient-based classical poisoning, QUID manipulates quantum encodings directly, requiring minimal retraining or optimization overhead. Its impact stems from exploiting quantum geometry rather than statistical gradients, making it a uniquely quantum-specific attack. To mitigate this, Q-Detection~\cite{he2025q} can be employed as a hybrid quantum-classical defense. We empirically evaluate QUID and Q-detection in section \ref{sec:quid-eval}.

\subsection{White-Box Attack}
\label{subsec:White-box}

In the white-box setting, adversaries have full visibility and control over the QML system, including its architecture, parameters, compilation artifacts, and hardware execution stack~\cite{ghosh2025ai,xu2025security,Ghosh2024Imitation,Saeed2019Locking,rehman2025opaque}. This is the strongest threat model, capturing malicious insiders, compromised service operators, or attackers with privileged access to the transpilation pipeline and backend hardware. Such adversaries can combine classical manipulations, such as adversarial optimization with quantum-specific mechanisms including pulse tampering and qubit remapping. 
This enables direct compromise of confidentiality, integrity, and availability, through parameter exfiltration, circuit modification, or disruption of quantum operations.


\paragraphB{Backdoor Attacks} 
In QML, an attacker can directly modify quantum circuits to embed trigger-dependent malicious functionality as demonstrated by QTrojan~\cite{chu2023qtrojan}, and exploit the training of hybrid classical-quantum neural networks to inject poisoned data and embed triggers~\cite{guo2025backdoor}.


\paragraphB{Reverse Engineering Attacks (REAs)}
In QML, REAs exploit transpiled, hardware-specific circuits to recover the original hardware-agnostic model, including its entanglement structure and trained parameters.
Prior work shows that gate-pattern analysis, lookup-table reconstruction, and parameter search can recover approximate duplicate of the target QNNs from transpilation artifacts~\cite{Ghosh2024Imitation}, while later approaches use autoencoder-based methods to reduce extraction cost while preserving high functional similarity~\cite{ghosh2025ai}. These results identify transpilation artifacts as a distinct leakage channel in QML and highlight model confidentiality as a central concern in cloud-based deployment.

\paragraphB{Hardware/ Compiler Trojan Attack}
Adversaries with privileged access to the transpilation toolchain or backend can implant malicious logic during compilation~\cite{roy2024hardwaretrojans}. By modifying coupling maps, inserting stealthy ancilla-controlled gates, or altering pulse schedules, they can create trojans that remain dormant in simulation but activate on hardware under specific conditions. These attacks can manipulate outputs, leak model parameters, and degrade hardware reliability, threatening confidentiality, integrity, and availability.

\paragraphB{Input Inference Attack}
An adversary with access to internal parameters and gradients can reconstruct sensitive inputs from training or inference~\cite{Heredge2025}. By exploiting the relationship between model parameters and encoded quantum states, the attacker can recover input features or their internal representations. This directly threatens confidentiality and may enable targeted downstream attacks.

\paragraphB{Adversarial Evasion Attacks}
With access to full model gradients and parameters, attackers can craft minimal perturbations to encoded inputs that cause targeted misclassification~\cite{lu2020quantum,west2023benchmarking,gong2022universal}. Methods such as FGSM and PGD \cite{west2023benchmarking,west2023towards} modify input encodings or amplitudes, exploiting sensitivity in the parameterized quantum circuit. These classical-style evasion attacks highlight the fragility of QML decision boundaries under white-box conditions. We evaluate FGSM and PGD white-box attacks on our QMLP model using angle-encoding and amplitude-encoding in Section~\ref{sec:PGD-FGSM}.

\subsection{Training vs. Inference-Time Decomposition}
\label{sec:attack-phase}

Beyond access level, a key complementary dimension is \emph{when} and attack occurs in the QML life-cycle. Training-time and inference-time attacks target different system artifacts, requiring distinct capabilities, and demand different defenses. Table\ref{tab:Attack_taxonomy} and figure\ref{fig:Taxonomy} maps each attack category in our taxonomy to its deployment phase.


\paragraphB{Training-Time Attacks (TTA)} 
TTA corrupt model development so that the trained model behaves incorrectly at deployment. Their effect is embedded in the learned parameters or decision boundaries before inference. In QML, such attacks can be especially damaging because training is costly. often requiring many epochs and quantum circuit executions, making retraining expensive. These attacks primarily target training data and compilation artifacts. Data poisoning attacks, including label-flipping~\cite{Bhatia2024QFL-LF} and QUID~\cite{Kundu-poisoning-2025}, alter labels to reduce accuracy or induce targeted misclassification. QUID is quantum-specific in that it exploits Encoder State Similarity (ESS) in Hilbert space to choose samples for relabeling, achieving substantially higher attack success than random corruption (Section~\ref{sec:quid-eval}). At the circuit level, hardware and compiler Trojan attacks~\cite{roy2024hardwaretrojans} implant malicious logic during transpilation that persists through training.

\paragraphB{Inference-Time Attacks (ITAs)} 
ITAs target deployed models without altering their parameters, instead manipulating inputs, hardware behavior, or outputs to induce errors or extract sensitive information.
Adversarial evasion is the most prominent inference-time threat. White-box methods such as FGSM and PGD~\cite{west2023towards,west2023benchmarking} use loss gradients to craft misclassifying inputs, while black-box universal adversarial perturbations (UAPs)~\cite{gong2022universal} exploit transferability across quantum classifiers.
Other attacks target the execution environment or model interface. Side-channel and pulse-level attacks~\cite{choudhury25crosstalk,xu2025security,Shubha2025Pulse} exploit multi-tenant NISQ hardware to leak metadata or induce errors during execution. Model extraction attacks~\cite{kundu2024evaluating,Fu2025CopyQNN,Fu2024QuantumLeak} query prediction interfaces to replicate model behavior, while reverse engineering attacks~\cite{Ghosh2024Imitation,ghosh2025ai} analyze transpiled circuits to recover architecture and parameters. Input inference attacks~\cite{Heredge2025} reconstruct sensitive inputs from internal model states.

\paragraphB{Dual-Phase Attacks} Backdoor attacks span both training and inference: the trigger is embedded during training, but activates only at inference. This pattern appears across black-box backdoors such as QUAP~\cite{zhao2025backdoor} and QuPT~\cite{Bhowmik2025QuPT}, gray-box attacks including QDoor~\cite{chu2023qdoor} and proxy-model poisoning~\cite{huang2023backdoor}, and white-box attacks such as QTrojan~\cite{chu2023qtrojan} and poisoned hybrid QNN training~\cite{guo2025backdoor}. Hardware/compiler Trojans~\cite{roy2024hardwaretrojans} similarly implant malicious logic during compilation that is triggered at inference. This dual-phase structure requires defenses at both stages, including training-time sanitization and verification, and inference-time validation and anomaly detection (Section~\ref{sec:QML_pipeline}).



\section{Experimental Settings}
\label{sec: Experimental Settings}

\paragraphS{Model Architectures}
We have implemented QMLP models to analyze robustness against adversarial attacks under varying quantum conditions. Each QMLP follows a hybrid quantum-classical design in which the classical component is implemented in PyTorch and the quantum circuit is simulated using PennyLane’s \texttt{default.qubit} backend~\cite{Pennylane}. To emulaterealistic hardware behavior, depolarizing noise with probability $p=0.01$ is applied to single- and two-qubit gates ($RX$, $RZ$, $X$, and $CX$) via the \texttt{Qiskit Aer} backend ~\cite{aleksandrowicz2019qiskit}.

The QMLP architecture remains consistent across experiments, varying in circuit depth--2, 5, 10, and 50 layers, and encoding strategy--angle and amplitude. All models are trained using Adam optimizer with learning rate 0.001, weight decay, and batch size of 64. 
Baseline models are trained for 30 epochs. For attack experiments, we use early stopping (patience=5) with a maximum of 30 epochs.
For comparison with CMLP, we train a CMLP of similar architecture with similar training configurations. This enables a controlled analysis of whether observed vulnerabilities arise from quantum effects or from general learning dynamics. We also evaluated all QMLP models under a depolarizing noise setting ($p = 0.01$) to assess their behavior under realistic NISQ-era constraints. Each experiment was run three times with different seeds, and we report the mean.


\paragraphS{Baselines and Metrics}
We define our \emph{baseline models} as those trained without adversarial interference. They are trained under noiseless conditions and evaluated in both noiseless and depolarized environments ($p=0.01$). All attack and defense results are reported relative to these baselines. Model performance is primarily evaluated using accuracy as the key metric. 
To assess robustness under attack, we use \emph{relative accuracy}, defined as the ratio between the accuracy under attack and the clean (no-attack) accuracy. For any model,

\[
\text{Relative Accuracy} = \frac{\text{Acc}_{\text{under-attack}}}{\text{Acc}_{\text{baseline}}}.
\]

Values close to $1$ for relative accuracy indicate strong robustness, whereas smaller values reflect greater performance degradation under adversarial manipulation. We also report Attack Success Rate (ASR) for QUID~\cite{Kundu-poisoning-2025}, the backdoor attack by Huang et al.~\cite{huang2023backdoor}, and QTrojan~\cite{chu2023qtrojan}, as these attacks were replicated from prior studies within our experimental setup, and ASR was among the metrics used in the original works to demonstrate attack effectiveness. We maintained the same model architecture across all experiments.

\paragraphS{Dataset}
\label{subsec:Dataset}
We use two multiclass datasets from distinct domains to evaluate adversarial robustness in vision and cybersecurity: MNIST~\cite{deng2012mnist}, the most widely used dataset in prior QML works~\cite{west2023benchmarking, west2023towards}, and AZ-Class~\cite{madar}, a twenty-three-class Android malware dataset based on behavioral features. This dual-domain setting enables cross-domain evaluation of QML robustness under heterogeneous feature distribution

Our QMLP employs a 9-qubit circuit, which constrains the input dimensionality.  Therefore, both datasets are reduced using \emph{Principal Component Analysis (PCA)}, following prior QML practices~\cite{biamonte2017quantum}. For \emph{angle encoding}, inputs are reduced to nine principal components (one per qubit), whereas for \emph{amplitude encoding}, inputs are compressed to 512 dimensions to match the circuit's Hilbert space. 
This preprocessing ensures that feature representations align with the circuit's encoding capacity and NISQ hardware constraints. We use the full datasets for baseline, label-flipping, FGSM, and PGD experiments, performing 10-class classification on MNIST and 23-class classification on AZ-Class. For all other attacks, we use a stratified subset with 700 samples per class, yielding 7,000 training samples for MNIST and 16,100 for AZ-Class, and a test set of 2,000 samples for each dataset. To ensure fair comparison, separate baseline models are trained on the same stratified subsets for these attacks. Both of these baseline models performances are reported in Table \ref{tab:baseline-qmlp}.

\paragraphS{Framework and Computational Resources}
All experiments are conducted in a hybrid \texttt{PennyLane}-\texttt{PyTorch}-\texttt{Qiskit Aer} framework on two HPC clusters at the University of Texas at El Paso. The PUNAKHA cluster, featuring NVIDIA H100 GPUs on 8-way DGX and 4-way HGX nodes, is used for GPU-accelerated noiseless simulations. The JAKAR cluster, with 72 nodes containing dual Intel Xeon Gold 6230 processors, is used for CPU-intensive noisy simulations. Noiseless runs use the \texttt{lightning.qubit} and \texttt{default.qubit} backends on PUNAKHA, while noisy runs use \texttt{default.mixed} and \texttt{Aer} on JAKAR due to the density-matrix formalism. To emulate NISQ behavior, we apply depolarizing noise (p=0.01) to single- and two-qubit gates to emulate NISQ behavior.

\section{Evaluation}

\begin{table}[!t]
\footnotesize
\caption{\textbf{Summary of Baseline Results.} Accuracies (\%) of QMLP across encoding schemes and circuit depths for MNIST and AZ-Class datasets under noiseless and depolarized noise ($p=0.01$) conditions. L.: Layers, Enc.: Encoding, Ang.: Angle, Amp.: Amplitude, NL = Noiseless, DN = Depolarized Noise. F.= Models Trained and Tested on Full Dataset, S.: Models Trained and Tested on Subset.}
\label{tab:baseline-qmlp}
\centering
\setlength{\tabcolsep}{3pt}
\begin{tabular}{p{0.55cm}|c|cc|cc|cc|cc}
\toprule
\multirow{3}{*}{\textbf{Enc.}} & 
\multirow{3}{*}{\textbf{L}} & 
\multicolumn{4}{c|}{\textbf{AZ-Class}} & 
\multicolumn{4}{c}{\textbf{MNIST}} \\ \cline{3-10}
 & & \multicolumn{2}{c|}{\textbf{NL}} & \multicolumn{2}{c|}{\textbf{DN}} & \multicolumn{2}{c|}{\textbf{NL}} & \multicolumn{2}{c}{\textbf{DN}} \\ \cline{3-10}
 & & \textbf{F.} & \textbf{S.} & \textbf{F.} & \textbf{S.} & \textbf{F.} & \textbf{S.} & \textbf{F.} & \textbf{S.} \\
\midrule
\multirow{4}{*}{\rotatebox[origin=c]{90}{Ang.}}
  & 2  & 49.8 & 46.0 & 47.0 & 46.1 & 66.0 & 41.9 & 59.2 & 41.7 \\
  & 5  & 52.3 & 47.1 & 22.4 & 46.5 & 68.0 & 53.9 & 45.0 & 52.0 \\
  & 10 & 54.8 & 48.7 & 4.9  & 41.8 & 83.6 & 55.8 & 23.6 & 48.2 \\
  & 50 & 32.4 & 28.6 & 5.2  & 21.3 & 76.9 & 52.4 & 9.9  & 32.1 \\
\midrule
\multirow{4}{*}{\rotatebox[origin=c]{90}{Amp.}}
  & 2  & 40.1 & 34.3 & 5.1  & 7.5  & 50.7 & 45.5 & 9.9  & 18.1 \\
  & 5  & 48.1 & 42.1 & 5.0  & 6.6  & 70.3 & 52.9 & 9.5  & 18.7 \\
  & 10 & 55.0 & 45.7 & 4.78 & 6.2  & 79.8 & 60.82& 10.1 & 16.4 \\
  & 50 & 67.0 & 55.7 & 4.81 & 7.6  & 92.6 & 84.3 & 9.7  & 15.5 \\
\midrule[\heavyrulewidth]
{\scriptsize CMLP} & -- & 95.89 & -- & -- & -- & 96.63 & -- & -- & -- \\
\bottomrule
\end{tabular}
\end{table}

\label{sec:5-evaluation}

This section aims to systematically examine how \emph{encoding schemes}, \emph{circuit depth}, and \emph{quantum noise} jointly influence model vulnerability, based on the dataset and experimental setup presented in the previous section, thereby addressing \protect\circled{RQ2}. We organize this section by evaluating each representative attack and its corresponding defense under controlled variations of encoding, circuit depth, and noise parameters, to address \protect\circled{RQ3}. For each threat model category (black-box, gray-box, and white-box), we analyze the impact of these factors on model robustness and discuss the resulting performance trends.

\subsection{Baseline}
\label{sec:QMLPBaseline}

\begin{tcolorbox}[
    colback=blue!5,
    colframe=blue!40!black,
    colbacktitle=blue!15!white,
    coltitle=black,
    fonttitle=\bfseries,
    title filled,
    boxrule=0.8pt,
    arc=2mm,
    left=4mm, right=4mm, top=2mm, bottom=2mm,
    title={{\em Key Insights: Baseline}}
]
\begin{itemize}
    \item Encoding and circuit depth jointly determine QMLP performance; angle encoding favors shallow circuits (highest accuracy gain for 10 layer PQC), while amplitude encoding benefits from deeper ones.
    \item Noise severely limits model accuracy, with shallow angle-encoded circuits showing relatively higher resilience.
\end{itemize}
\end{tcolorbox}

We report baseline performance for QMLP models under noiseless and noisy conditions , compared against a CMLP baseline. We analyze how encoding choice, circuit depth, and quantum noise affect accuracy and stability across datasets.

\begin{figure*}[!t]
    \centering

    \begin{subfigure}[b]{0.24\textwidth}
        \centering
        \includegraphics[width=\textwidth]{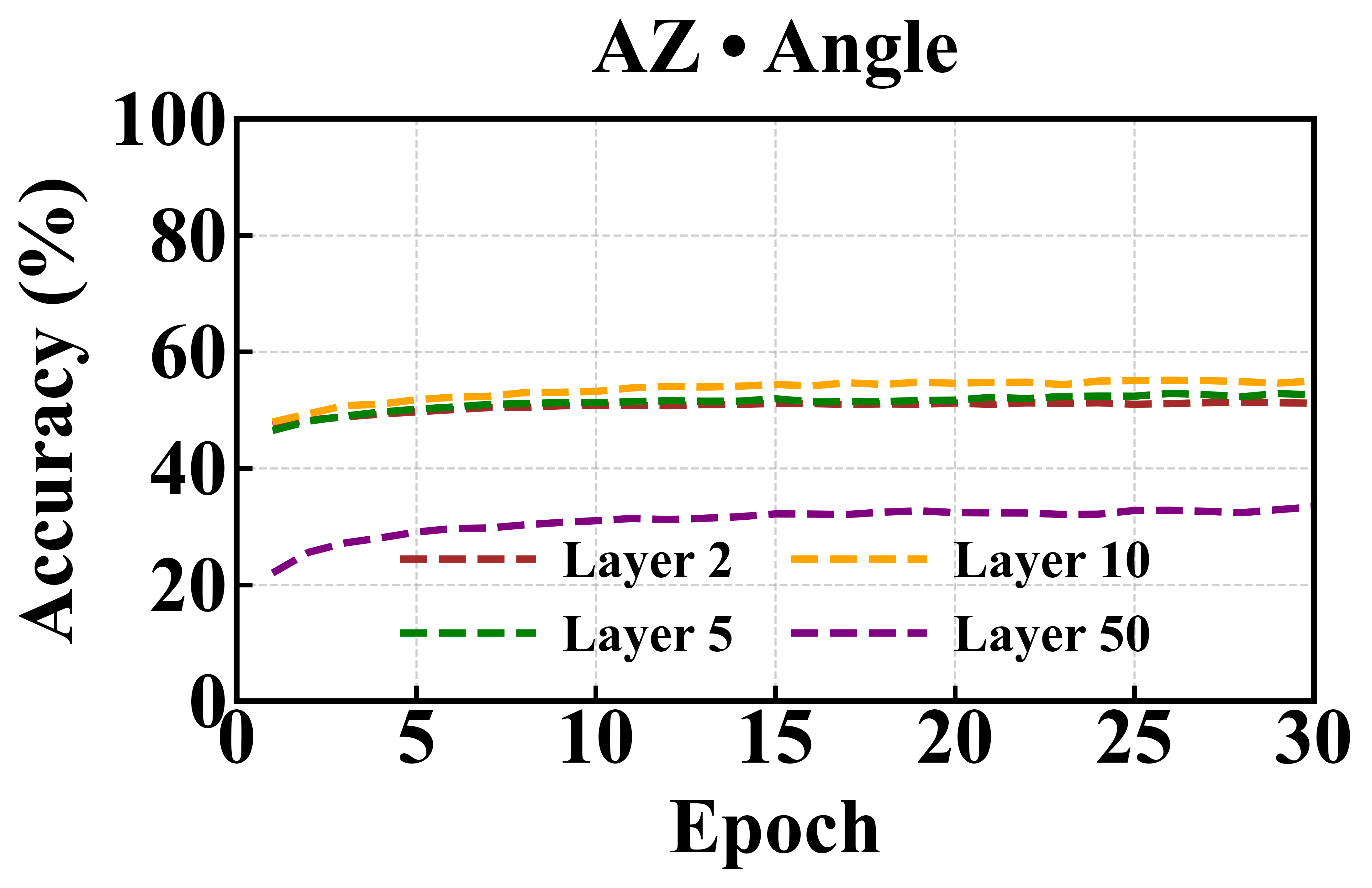}
        \label{fig:QMLP_baseline_az_angle}
    \end{subfigure}
    \hfill
    \begin{subfigure}[b]{0.24\textwidth}
        \centering
        \includegraphics[width=\textwidth]{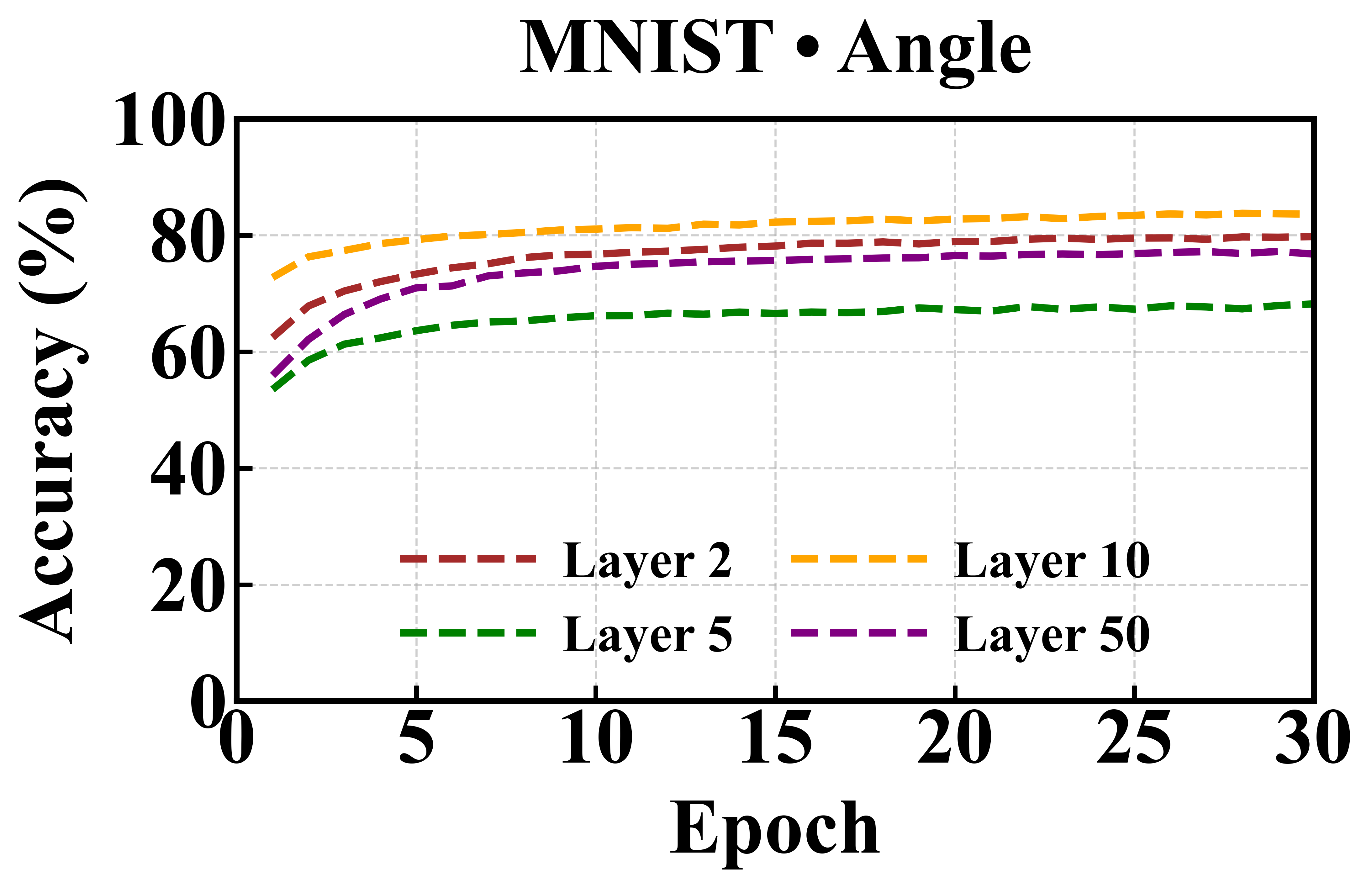}
        \label{fig:QMLP_baseline_az_amplitude}
    \end{subfigure}
    \hfill
    \begin{subfigure}[b]{0.24\textwidth}
        \centering
        \includegraphics[width=\textwidth]{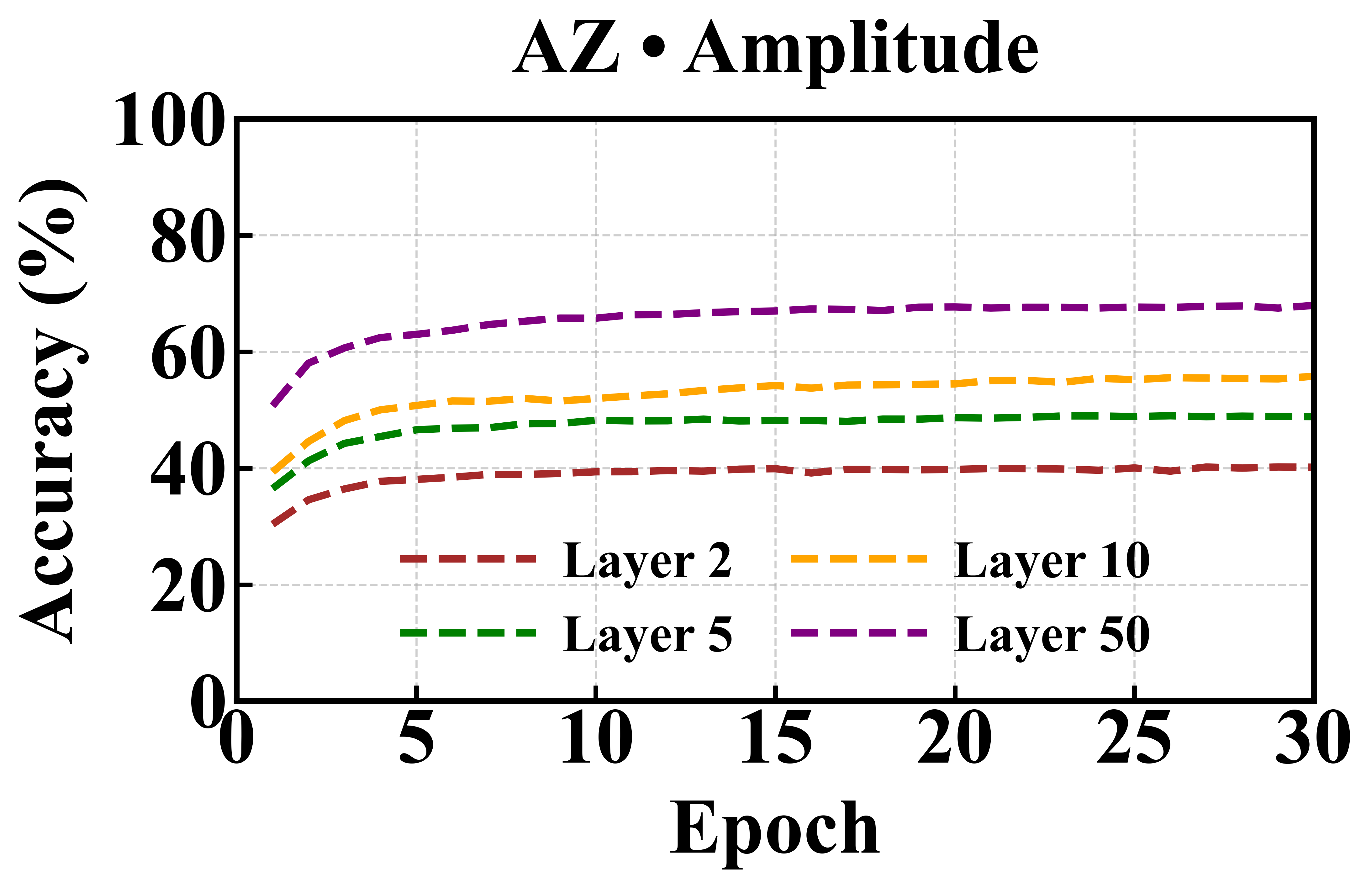}
        \label{fig:QMLP_baseline_MNIST_angle}
    \end{subfigure}
    \hfill
    \begin{subfigure}[b]{0.24\textwidth}
        \centering
        \includegraphics[width=\textwidth]{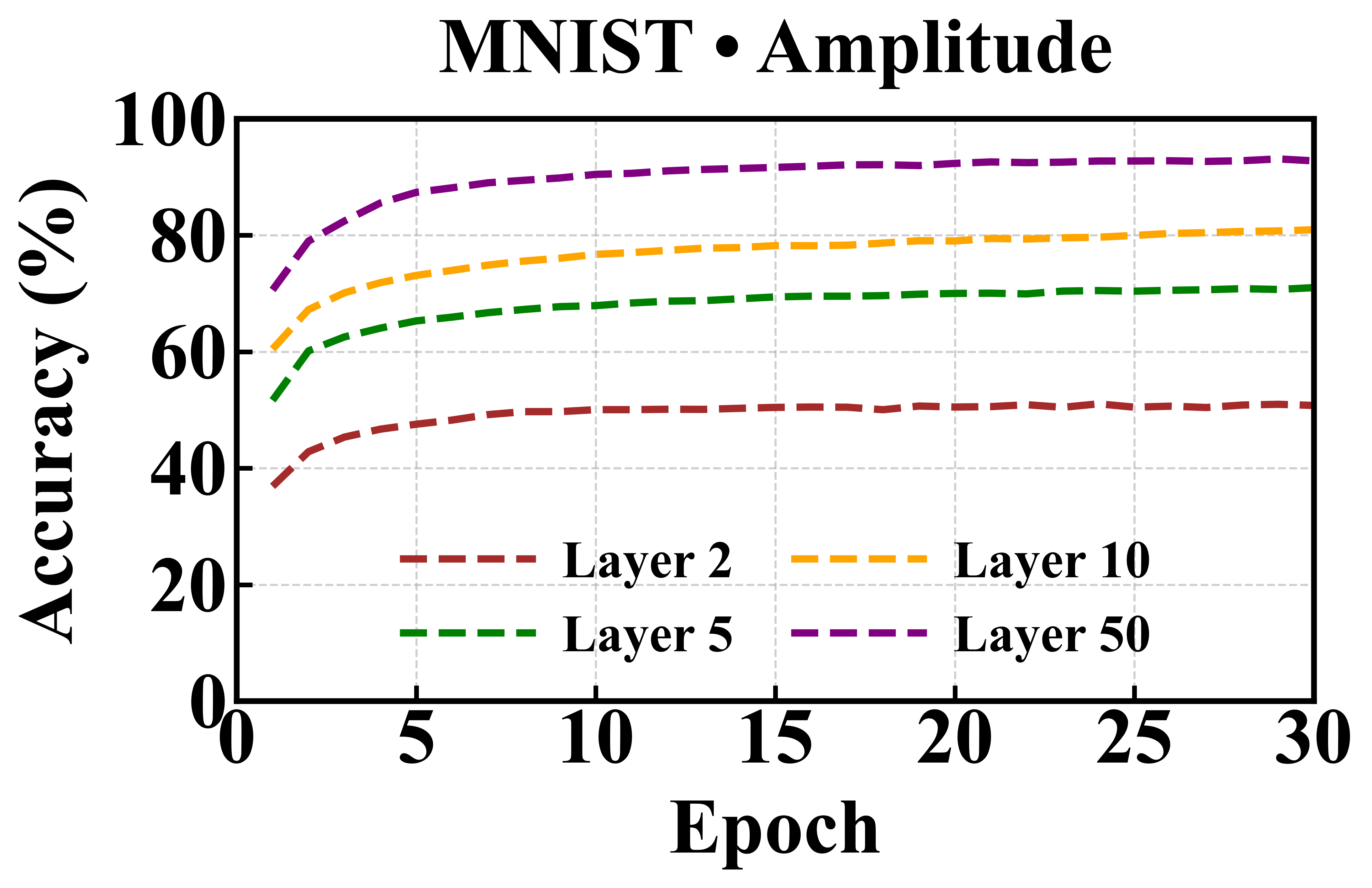}
        \label{fig:QMLP_baseline_MNIST_amplitude}
    \end{subfigure}
    \vspace{-0.3cm}
    \caption{
    Performance of the QMLP model with varying circuit layers and encoding schemes on the AZ-Class and MNIST datasets under noiseless condition. 
    }
    \label{fig:QMLP_baseline}
\end{figure*}

\paragraphS{QMLP under Noiseless Setting}
\label{sec:QMLPNoiseless}
Table~\ref{tab:baseline-qmlp} presents the baseline performance of the QMLP models, while Figure~\ref{fig:QMLP_baseline} illustrates their accuracy progression over epochs.
At shallow depths (2 layers), angle encoding performs better, achieving 66\% accuracy on MNIST and 49.8\% on AZ-Class. This advantage stems from its localized qubit-wise mapping, which remains stable when entanglement is minimal. However, as circuit depth increases, accuracy improves only up to 10 layers. Beyond ten layers, angle encoding degrades significantly-dropping to 32.4\% on AZ-Class and 76.9\% on MNIST at 50 layers-likely due to destructive interference and over-entanglement that hinder convergence. For angle encoding, 10 layer models achieve the highest accuracy in both the datasets. In contrast, amplitude encoding consistently improves with depth. Accuracy rises from 50.7\% at 2 layers to 92.6\% at 50 layers on MNIST and from 40.1\% to 67\% on AZ-Class. This pattern indicates that deeper amplitude-encoded circuits leverage the Hilbert-space representation more efficiently.

In summary, angle encoding gains modestly upto a certain depth, approximately 5-20\%, while amplitude encoding improves dramatically, with up to 42\% gains on MNIST and 27\% on AZ-Class. These findings establish that {\em circuit depth is far more influential for amplitude-encoded models than for angle-encoded ones}.

\paragraphS{QMLP under Depolarized Noise Setting (\(p = 0.01\))}
Table~\ref{tab:baseline-qmlp} also reports results under depolarizing noise, emulating NISQ-era conditions.
For angle encoding, the 2-layer QMLP reaches 59.2\% accuracy on MNIST, but performance declines to 9.9\% at 50 layers as cumulative noise overwhelms coherence. For amplitude encoding, accuracy remains near 10\% across all depths, indicating high noise sensitivity. A similar pattern appears in AZ-Class, where both encodings perform substantially worse than on MNIST.

\paragraphS{Classical Multi-layer Perceptron (CMLP)}
The CMLP baseline achieves 96.63\% test accuracy on MNIST and 95.89\% on AZ-Class (Table~\ref{tab:baseline-qmlp}). These results define upper performance bounds and underscore the gap between mature classical networks and current QML models. Unlike QMLPs, CMLPs are unaffected by quantum noise or circuit-depth limitations, highlighting that current NISQ constraints, not learning inefficiency, are the dominant cause of reduced quantum performance.

\paragraphS{QMLP Vs CMLP Comparison}
The performance gap between QMLP and CMLP arises mainly from hardware-driven dimensionality constraints rather than learning inefficiency. Current NISQ devices typically support only about 8--10 practical qubits, requiring substantial PCA-based dimensionality reduction before quantum encoding. To remain representative of real hardware, QMLP simulations must follow the same qubit and circuit constraints. As a result, angle encoding reduces inputs to nine principal components, while amplitude encoding reduces them to 512 dimensions, discarding much of the original feature information. By contrast, CMLP models use the full unreduced feature space and retain all discriminative information. This imbalance in input representation, rather than a fundamental weakness of quantum learning, is the primary reason for the observed accuracy gap between QMLP and CMLP.

\begin{table}[!t]

\caption{\textbf{Label-Flipping Attack} 
\label{tab:label-flipping_ratio}
Results are reported as \emph{accuracy ratios} under label-flipping attack, relative to clean baselines.
\textbf{N} = No Label Smoothing, \textbf{Y} = With Label Smoothing.
\textbf{CMLP} baselines: 96.63\% (MNIST), 95.89\% (AZ-Class).
\textbf{QMLP-Angle} and \textbf{QMLP-Amplitude} are Quantum MLPs using angle and amplitude encodings, respectively.
\textbf{AccNL.} = Noiseless; 
\textbf{AccDN.} = Depolarized-Noise}
\centering
\scriptsize
\renewcommand{\arraystretch}{1.10}
\setlength{\tabcolsep}{3pt}

\begin{tabular}{@{}c|c|c|cc|cc@{}}
\toprule
\multirow{2}{*}{\textbf{Model}} &
\multirow{2}{*}{\textbf{Layers}} &
\multirow{2}{*}{\textbf{LS}} &
\multicolumn{2}{c|}{\textbf{AZ-Class (ratio)}} &
\multicolumn{2}{c}{\textbf{MNIST (ratio)}} \\ 
\cline{4-7}
 &  &  & \textbf{AccNL.} & \textbf{AccDN.} & \textbf{AccNL.} & \textbf{AccDN.} \\
\midrule

\multirow{2}{*}{CMLP}
 & -- & N & 0.51 & -- & 0.50 & -- \\
 & -- & Y & 0.59 & -- & 0.51 & -- \\
\midrule

\multirow{6}{*}{QMLP-Angle}
  & 2  & N & 0.93 & 0.95 & 0.93 & 0.89 \\
  & 5  & N & 0.91 & 1.73 & 0.95 & 1.00 \\
  & 10 & N & 0.91 & 3.33 & 0.79 & 0.73 \\ \cline{2-7}
  & 2  & Y & 0.90 & 0.87 & 0.92 & 0.82 \\
  & 5  & Y & 0.91 & 0.74 & 0.95 & 0.63 \\
  & 10 & Y & 0.89 & 2.00 & 0.79 & 0.55 \\
\midrule

\multirow{6}{*}{QMLP-Amplitude}
  & 2  & N & 0.94 & 1.10 & 1.03 & 0.98 \\
  & 10 & N & 0.92 & 1.03 & 0.95 & 0.95 \\
  & 50 & N & 0.93 & 1.00 & 0.97 & 1.02 \\ \cline{2-7}
  & 2  & Y & 0.87 & 1.02 & 1.06 & 1.01 \\
  & 10 & Y & 0.91 & 1.05 & 0.98 & 0.99 \\
  & 50 & Y & 0.90 & 0.98 & 0.97 & 1.05 \\

\bottomrule
\end{tabular}
\end{table}

\paragraphS{Statistical Analysis}
We report standard errors, 95\% confidence intervals via the $t$-distribution, and conduct $t$-tests across key comparisons. Standard errors remain below $\pm 3$ percentage points across all configurations. The accuracy drop from noiseless to depolarizing noise is statistically significant across all encodings and depths (paired $t$-test, $p{=}0.004{<}0.05$ for the least significant case). At shallow depth, angle encoding significantly outperforms amplitude encoding (Welch's $t$-test, $p{=}0.002{<}0.05$ on AZ-Class), whereas at 10 layers the difference is not significant ($p{=}0.89{>}0.05$). Increasing depth significantly improves noiseless accuracy for both encodings ($p{=}0.019{<}0.05$ for angle; $p{<}0.0001$ for amplitude). Under noise, shallow angle circuits significantly outperform deeper ones ($p{<}0.0001$), while amplitude models collapse uniformly with no significant depth effect ($p{=}0.83{>}0.05$).

\subsection{Black-box $\rightarrow$ Data Poisoning (Label-Flipping)}
\label{sec:label-flipping}

\begin{tcolorbox}[
    colback=blue!5,
    colframe=blue!40!black,
    colbacktitle=blue!15!white,
    coltitle=black,
    fonttitle=\bfseries,
    title filled,
    boxrule=0.8pt,
    arc=2mm,
    left=4mm, right=4mm, top=2mm, bottom=2mm,
    title={\em Key Insights: Label-Flipping Attack}
]
\begin{itemize}

    \item QMLP models exhibit greater robustness than CMLP models under label-flipping attacks, driven partly by lower baseline accuracy and NISQ-imposed dimensionality reduction rather than fundamental quantum resistance.

    \item Label smoothing provides no meaningful benefit for QMLP and only modest gains for CMLP, as a 50\% poison ratio is too severe for soft-label regularization to counteract.
    
    \item Under depolarizing noise, relative accuracy may exceed 1, not due to true robustness, but because noise disrupts the poisoning signal when the clean baseline is already near-random.

\end{itemize}
\end{tcolorbox}

\begin{figure*}[!t]
    \centering
    \begin{subfigure}[b]{0.24\textwidth}
        \centering
        \includegraphics[width=\linewidth]{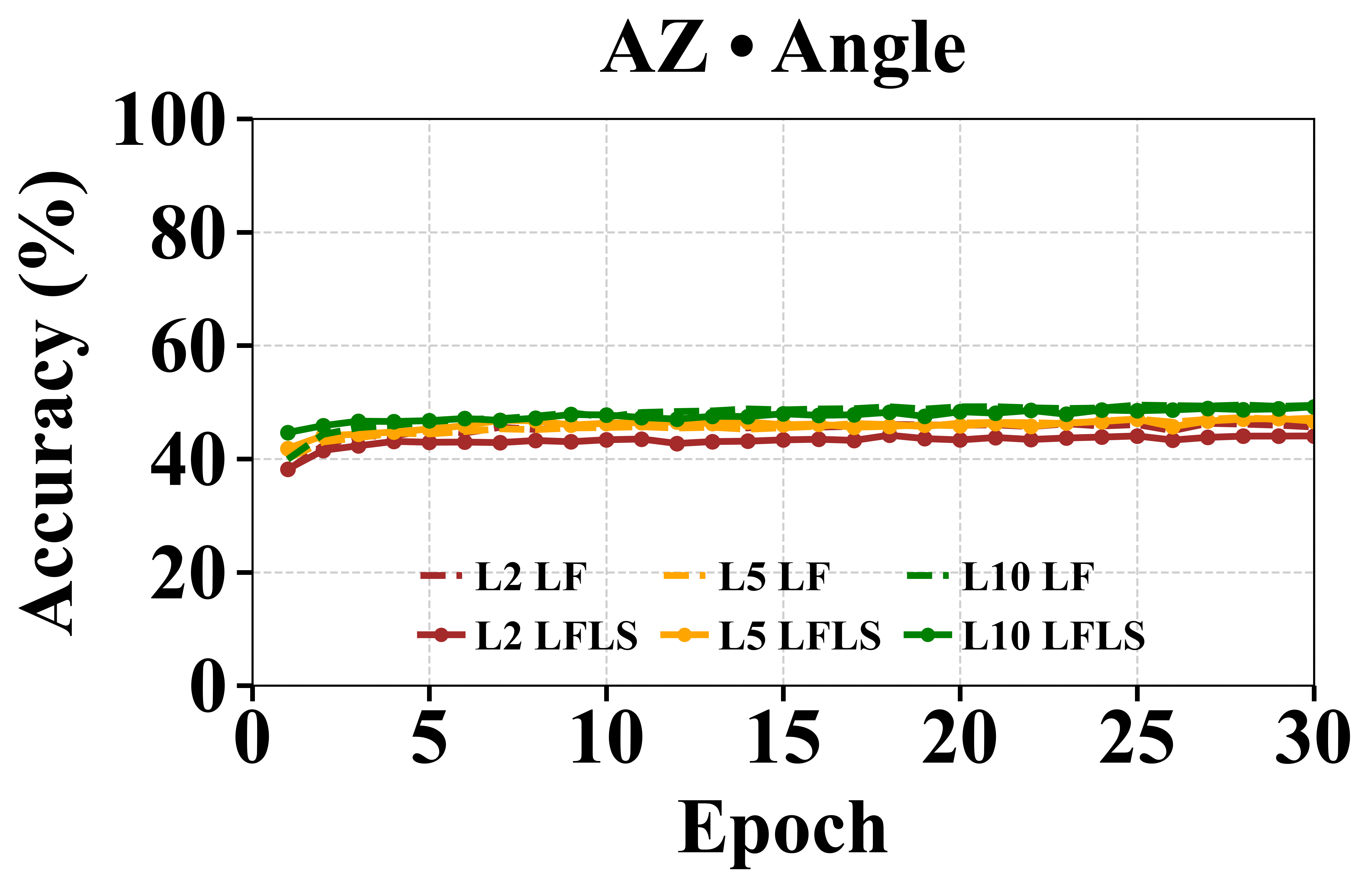}
        
    \end{subfigure}%
    \hfill%
    \begin{subfigure}[b]{0.24\textwidth}
        \centering
        \includegraphics[width=\linewidth]{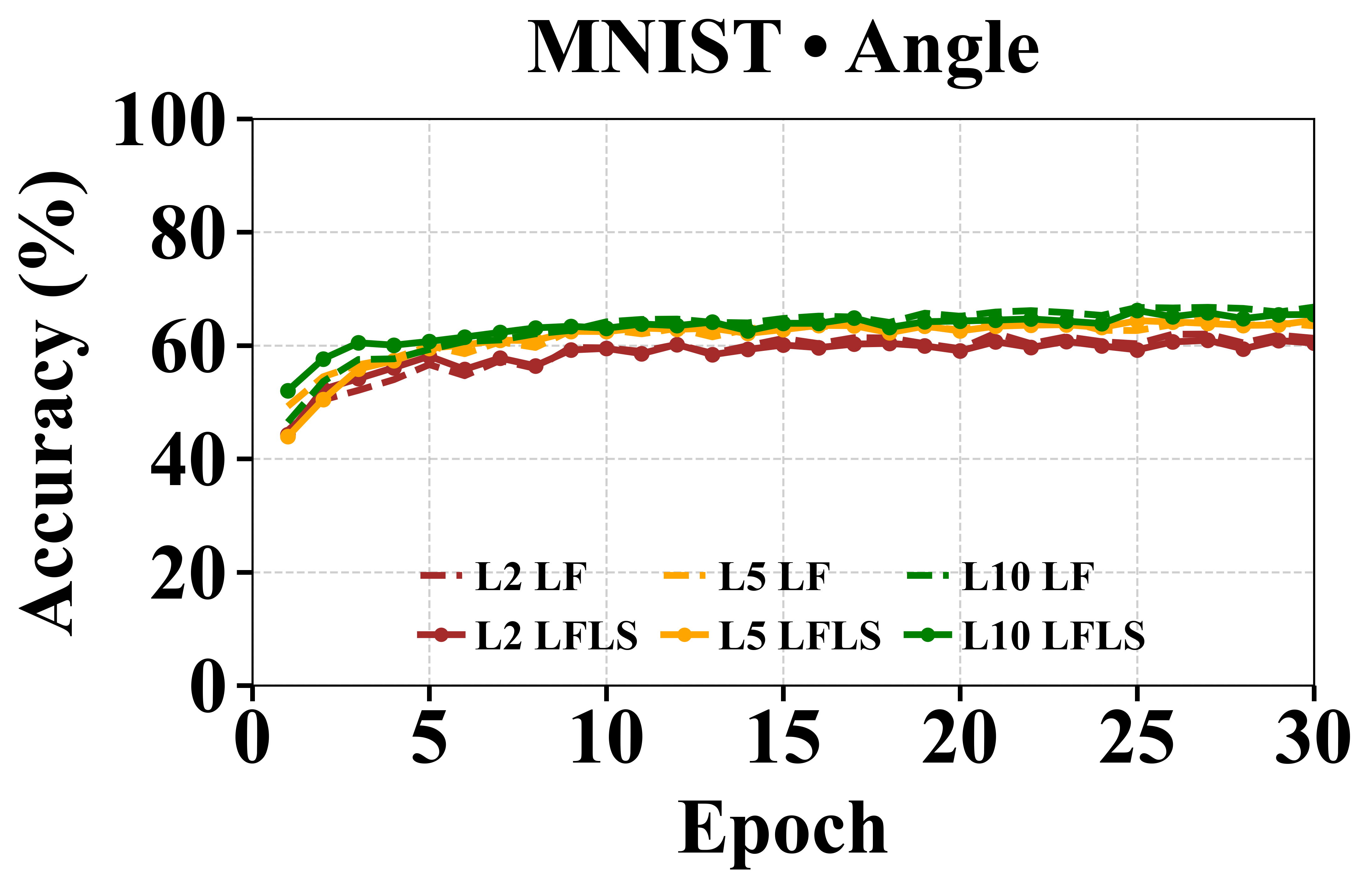}
       
    \end{subfigure}%
    \hfill%
    \begin{subfigure}[b]{0.24\textwidth}
        \centering
        \includegraphics[width=\linewidth]{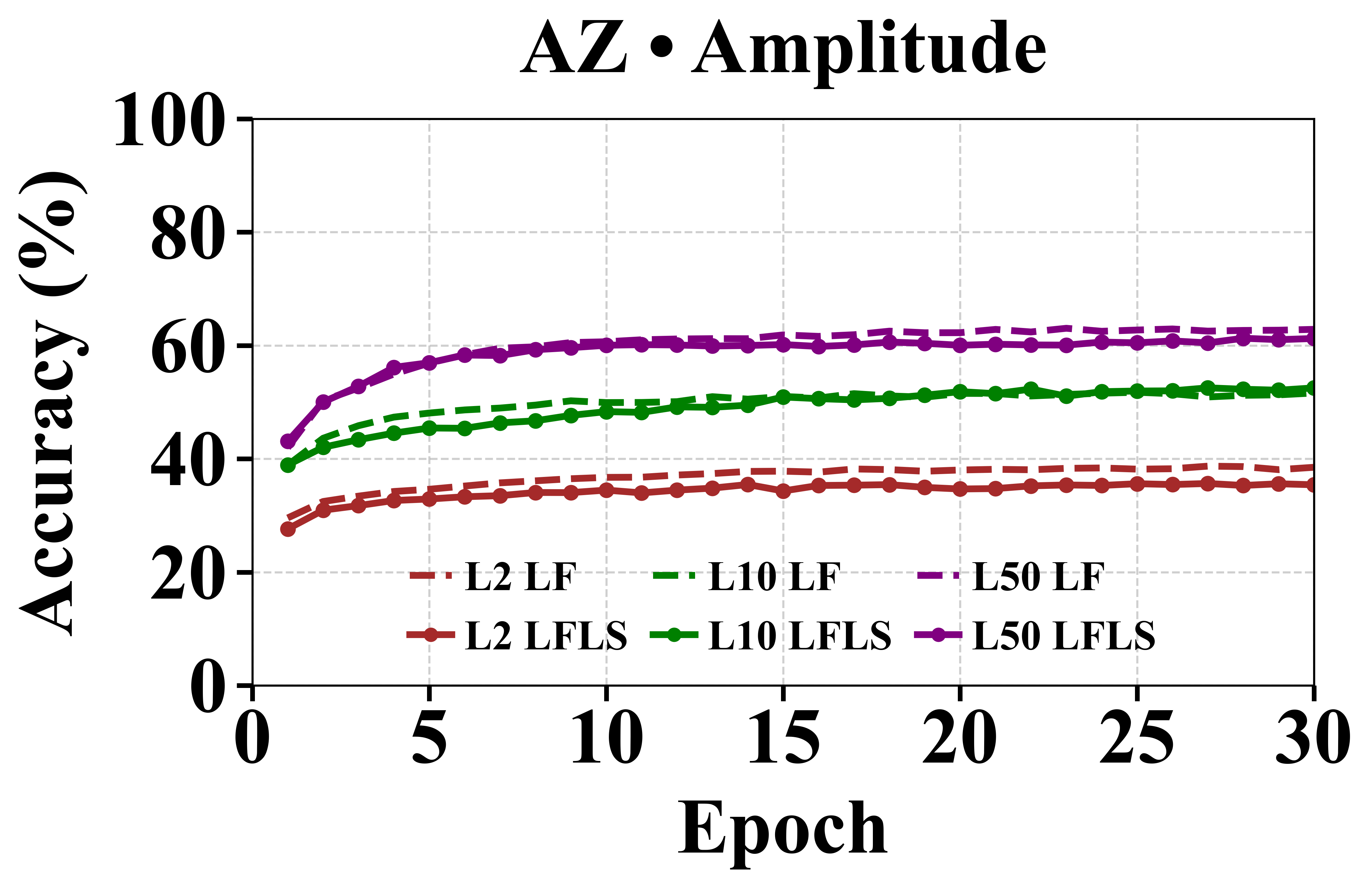}
       
    \end{subfigure}%
    \hfill%
    \begin{subfigure}[b]{0.24\textwidth}
        \centering
        \includegraphics[width=\linewidth]{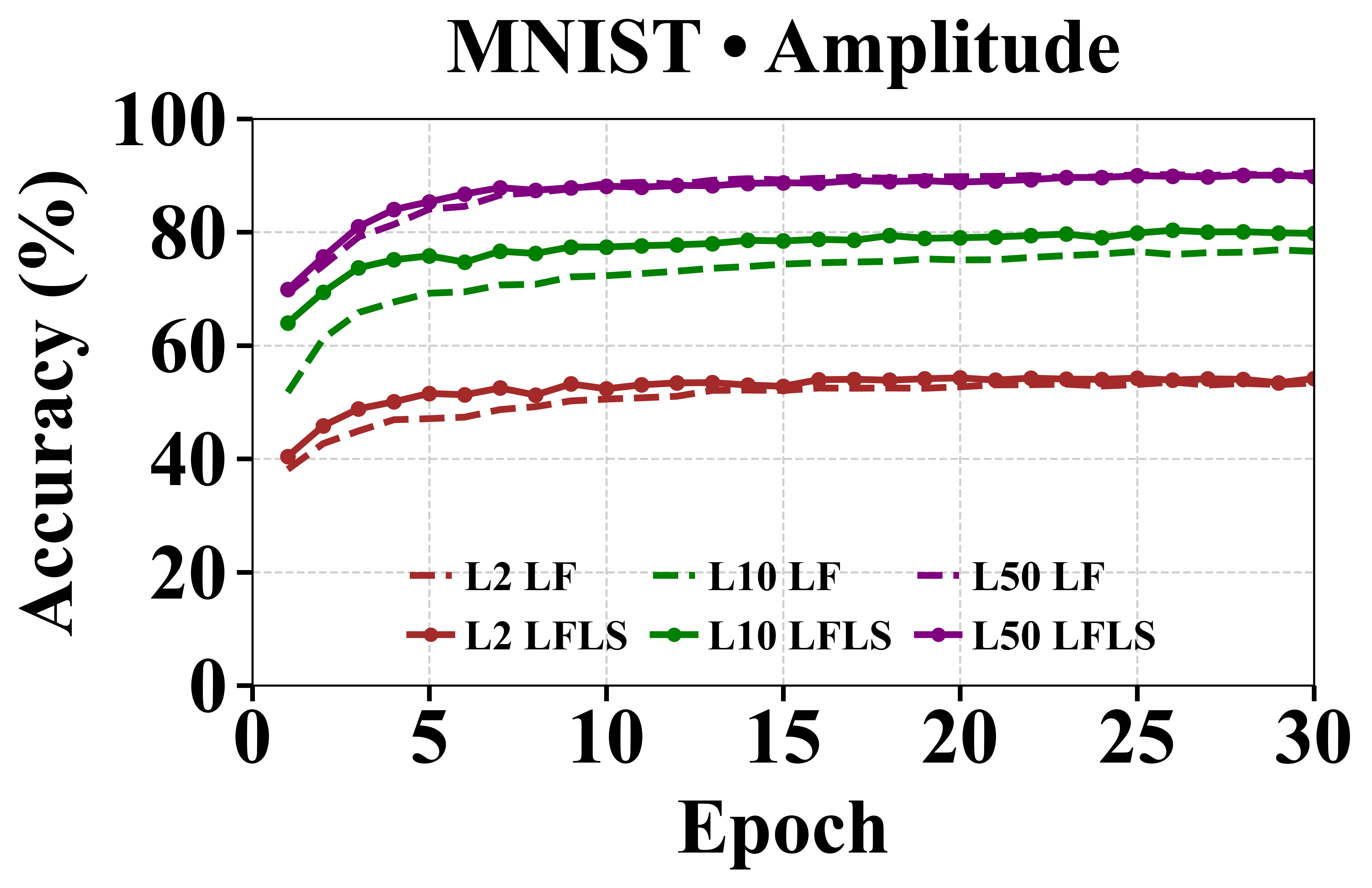}
       
    \end{subfigure}
    \vspace{-0.2cm}
    \caption{Performance of QMLP under label-flipping (LF) and label-flipping with label-smoothing (LFLS) across datasets (AZ-Class, MNIST) and encodings (Angle, Amplitude) in noiseless conditions.}
    \label{fig:Label-Flipping}
\end{figure*}

This attack models a training-time setting in which an adversary flips sample labels while leaving features unchanged, with the goal of degrading model accuracy. We implement a 50\% untargeted label-flipping attack on QMLP models with angle and amplitude encodings under noiseless training. For angle encoding, we evaluate 2-, 5-, and 10-layer circuits; for amplitude encoding, 2-, 10-, and 50-layer circuits, since 50-layer angle-encoded models are computationally infeasible. All models are trained noiselessly and tested under both noiseless and depolarizing noise (\(p=0.01\)), as reported in Table~\ref{tab:label-flipping_ratio}. To mitigate the attack, we adopt \emph{label smoothing}~\cite{Szegedy2015RethinkingTI} with \(\alpha=0.2\), which reduces overconfidence in corrupted labels. Figure~\ref{fig:Label-Flipping} shows test performance under noiseless conditions, with and without label smoothing.

\paragraphS{Performance under Noiseless Setting} Both angle- and amplitude-encoded QMLP models retain relative accuracies close to 90\% under a 50\% label-flipping attack, while the CMLP loses nearly 50\% of its clean accuracy. Two reasons explain this gap. First, a \emph{baseline ceiling effect}: the CMLP's high clean accuracy (96.63\% on MNIST, 95.89\% on AZ-Class) gives corrupted gradients a large margin to degrade, whereas QMLP baselines are far lower (e.g., 66\% on MNIST, 49.8\% on AZ-Class for 2-layer angle encoding, Table~\ref{tab:baseline-qmlp}), so the absolute accuracy drop is smaller and the relative ratio stays high. Second, NISQ-imposed PCA reduction- to 9 components for angle encoding and 512 for amplitude- discards much of the feature variation that flipped labels exploit, acting as inadvertent regularization unavailable to CMLP.

Label smoothing yields only a modest gain for CMLP (0.51 to 0.59 on MNIST) and no meaningful gain for QMLP. At $\varepsilon{=}0.5$, the corruption is too pervasive for smoothing with $\alpha{=}0.2$ to counteract; it would be more effective at lower poison ratios where overconfidence on a small number of corrupted labels is the dominant issue. For QMLP, the intervention finds no foothold regardless, since the model's already-low baseline accuracy means it does not operate in the overconfident regime that label smoothing targets.

\paragraphS{Performance under Depolarizing Noise}
Under depolarizing noise, angle-encoded QMLPs retain relative accuracies above 90\% in most settings. Some ratios exceed 1 (e.g., 1.73 for the 5-layer angle model on AZ-Class), but this does not indicate true robustness. In this case, the noisy clean baseline is already near-random (22.4\%, Table~\ref{tab:baseline-qmlp}), and depolarizing noise disrupts the coherent states on which the poisoning signal relies, causing the two effects to partially cancel rather than compound. Thus, despite high relative ratios, absolute accuracy drops sharply across configurations. Under noisy conditions, depolarizing noise, rather than label flipping, is the dominant source of degradation.

\paragraphS{Statistical Analysis}
Standard errors remain below $\pm 3$ percentage points across all configurations. The degradation from clean baseline to the label-flipping condition is significant across all QMLP configurations (paired $t$-test, $p{=}0.0013{<}0.05$ for the least significant case). The robustness gap between QMLP and CMLP under label flipping is statistically significant (Welch's $t$-test, $p{=}0.0001{<}0.05$). The further accuracy drop from noiseless to depolarizing noise evaluation is also significant ($p{=}0.0003{<}0.05$). Label smoothing yields a significant improvement for CMLP ($p{=}0.0102{<}0.05$). For QMLP, while the difference between LF and LFLS conditions reaches statistical significance in some configurations ($p{=}0.038{<}0.05$), but the absolute improvement is practically negligible.

\subsection{Gray-box $\rightarrow$ Quantum Indiscriminate Data (QUID) Poisoning Attack}
\label{sec:quid-eval}

\begin{figure*}[!ht]
    \centering

    \begin{subfigure}[b]{0.24\textwidth}
        \centering
        \includegraphics[width=\textwidth]{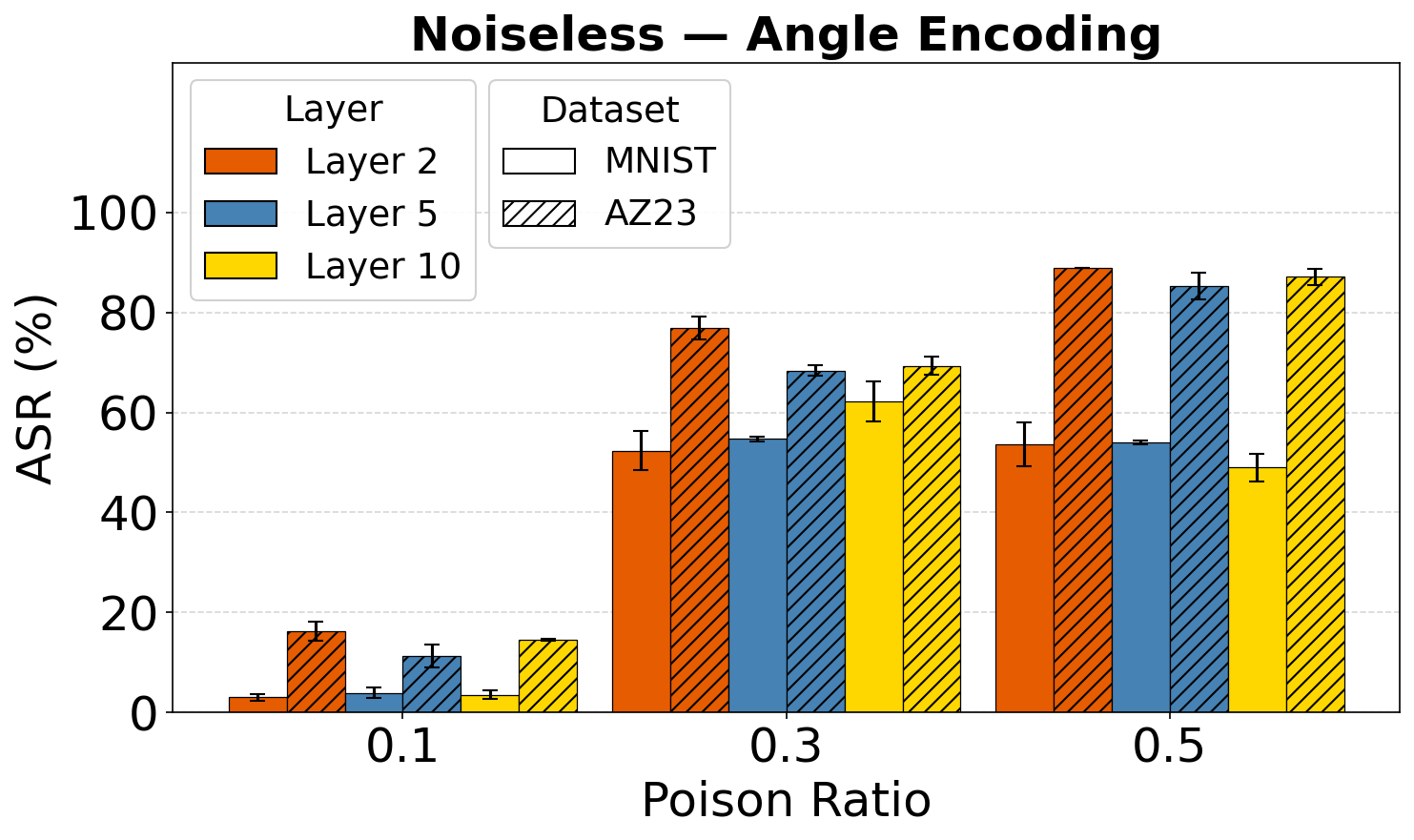}
    \end{subfigure}
    \hfill
    \begin{subfigure}[b]{0.24\textwidth}
        \centering
        \includegraphics[width=\textwidth]{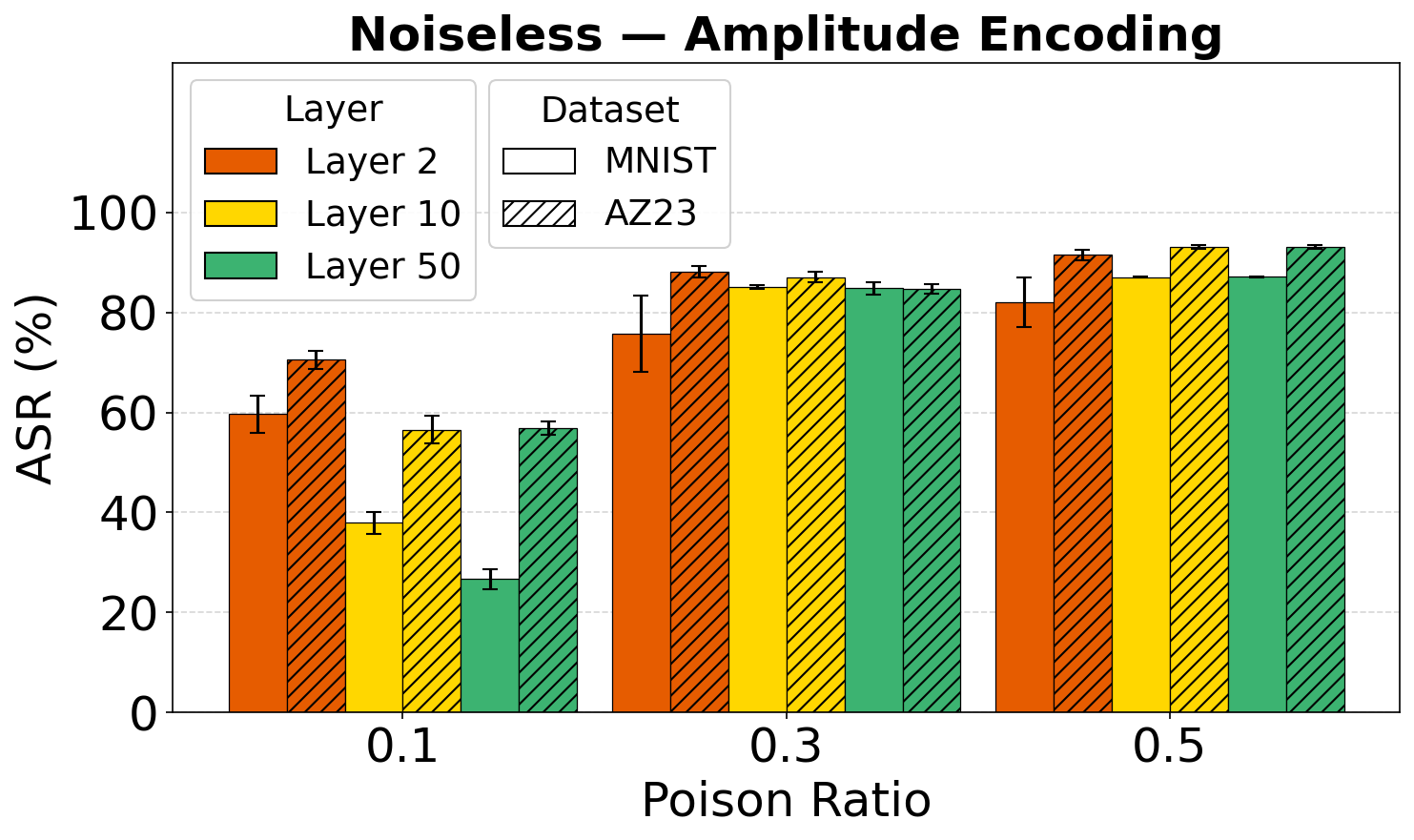}
    \end{subfigure}
    \hfill
    \begin{subfigure}[b]{0.24\textwidth}
        \centering
    
        \includegraphics[width=\textwidth]{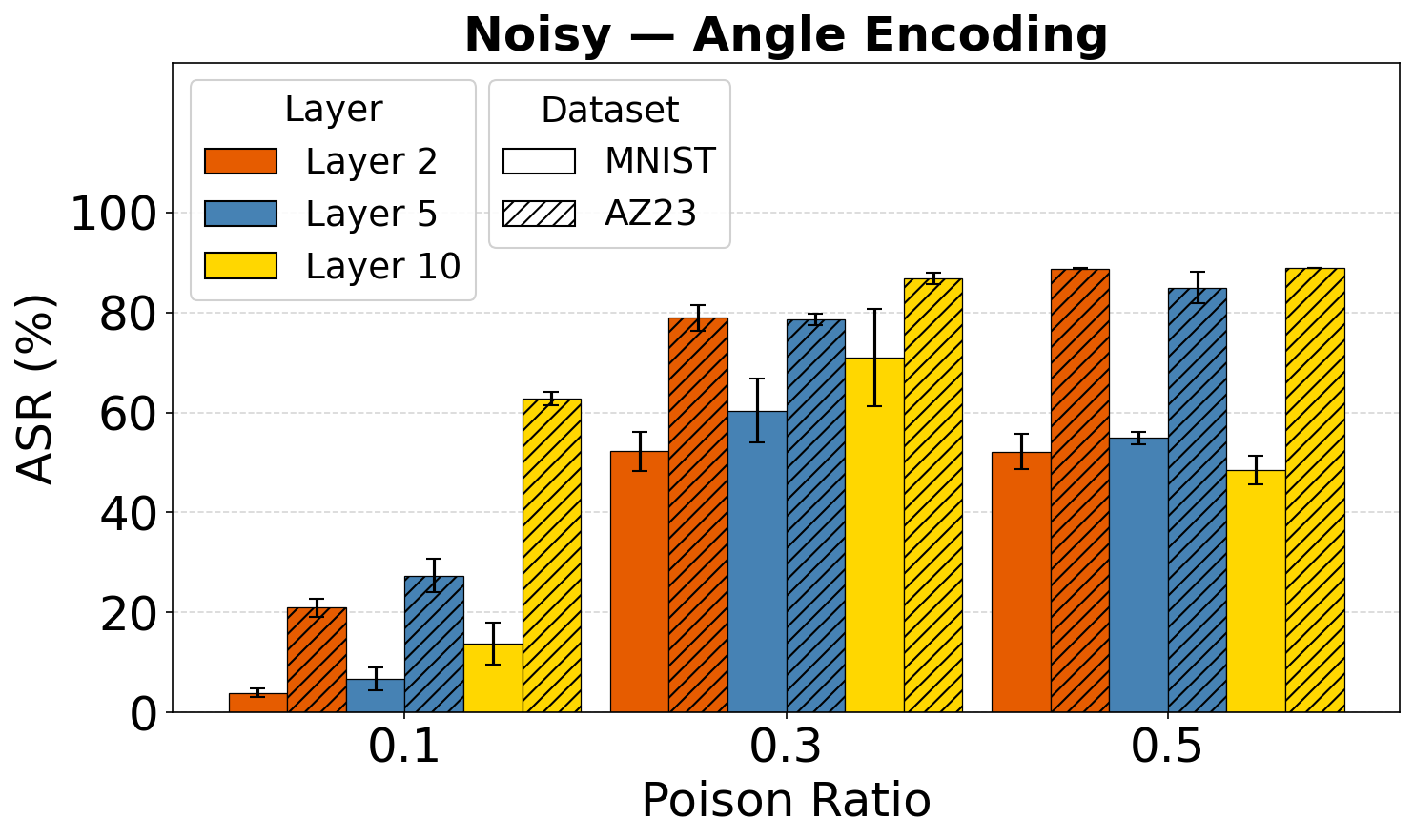}
    \end{subfigure}
    \hfill
    \begin{subfigure}[b]{0.24\textwidth}
        \centering
        \includegraphics[width=\textwidth]{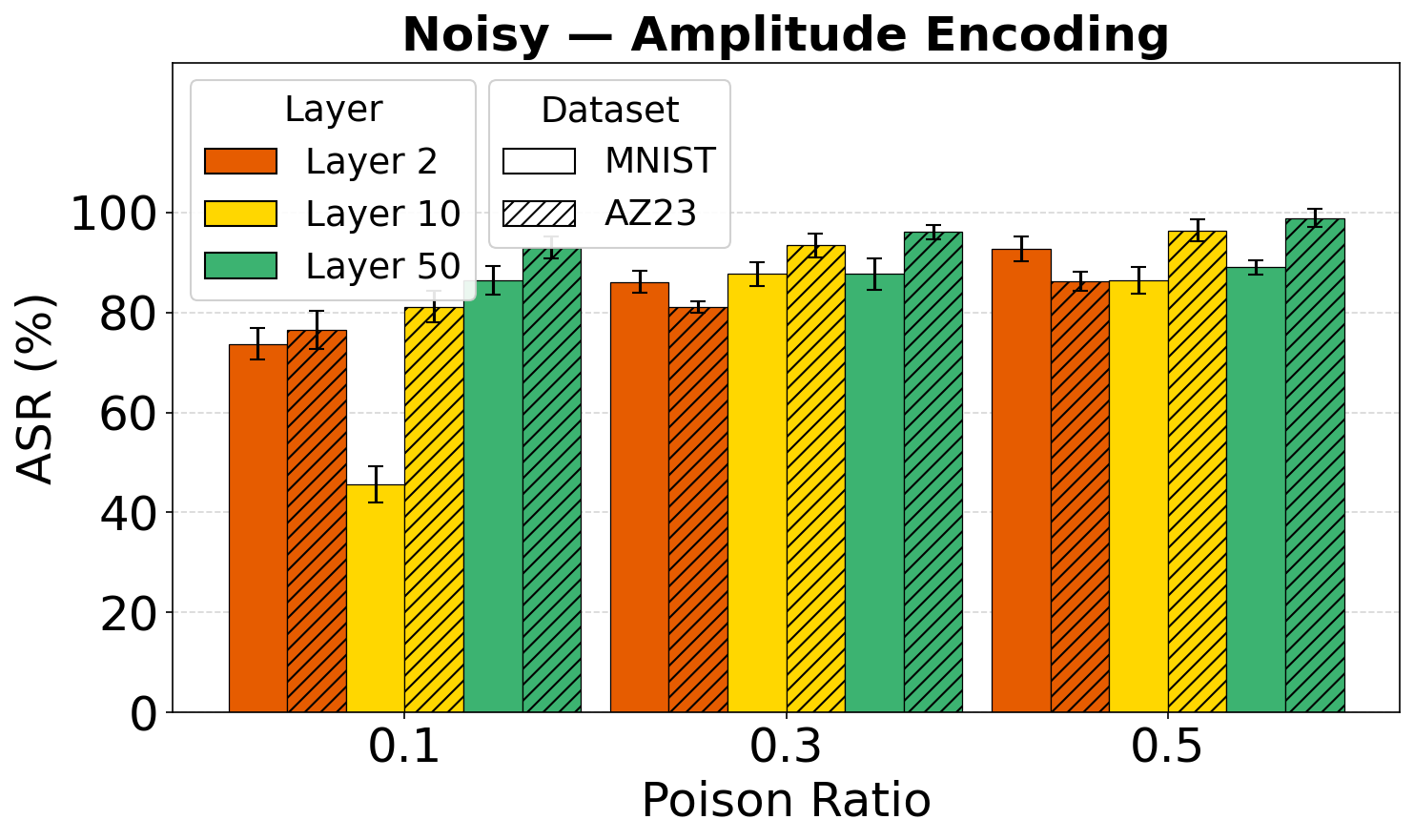}
    \end{subfigure}

    \caption{Attack Success Rate (ASR) of the QUID backdoor attack without Q-Detection across Amplitude and Angle encodings under Noiseless and Noisy quantum circuit settings with varying layer. 
    }
    \label{fig:quid-asr}
\end{figure*}

\begin{tcolorbox}[
colback=blue!5,
colframe=blue!40!black,
colbacktitle=blue!15!white,
coltitle=black,
fonttitle=\bfseries,
title filled,
boxrule=0.8pt,
arc=2mm,
title={\em Key Insights: QUID and Q-Detection}
]
\begin{itemize}
    \item Amplitude-encoded models are more vulnerable to QUID.
    \item Angle encoding is more resilient at low poison ratios but degrades nearly completely at $\varepsilon = 0.5$ across 
    all depths.
    \item Depolarizing noise weakens QUID on amplitude encoding by disrupting Hilbert-space structure, acting as a natural defense, while angle encoding remains largely vulnerable.
    \item Q-Detection is more effective with angle encoding at moderate poison ratios but offers limited benefit under noise.
    \item Even partial circuit access through the encoding scheme significantly compromises model integrity across all configurations.
\end{itemize}
\end{tcolorbox}

\begin{table}[!t]
\centering
\caption{\textbf{QUID Relative Accuracy (Ratio)} across encodings, layers, and poison ratios under noiseless (NL) and depolarized noise (DN, $p=0.01$) conditions, with and without Q-Detection (w/, w/o).}
\label{tab:quid_accuracy_ratio}
\footnotesize
\setlength{\tabcolsep}{2.5pt}
\renewcommand{\arraystretch}{1.1}
\begin{tabular}{@{}c|c|c|cc|cc|cc|cc@{}}
\toprule
\multirow{3}{*}{\textbf{Enc.}} & \multirow{3}{*}{\textbf{L}} & \multirow{3}{*}{\shortstack{\textbf{Poison}\\\textbf{Ratio}}} & \multicolumn{4}{c|}{\textbf{AZ-Class (Rel. Acc)}} & \multicolumn{4}{c}{\textbf{MNIST (Rel. Acc)}} \\ \cline{4-11}
& & & \multicolumn{2}{c|}{\textbf{NL}} & \multicolumn{2}{c|}{\textbf{DN}} & \multicolumn{2}{c|}{\textbf{NL}} & \multicolumn{2}{c}{\textbf{DN}} \\ \cline{4-11}
& & & \textbf{w/o} & \textbf{w/} & \textbf{w/o} & \textbf{w/} & \textbf{w/o} & \textbf{w/} & \textbf{w/o} & \textbf{w/} \\
\midrule
\multirow{9}{*}{\rotatebox[origin=c]{90}{\textbf{Amp.}}}
 & \multirow{3}{*}{2}  & 0.1 & 0.49 & 0.57 & 2.15 & 2.56 & 0.56 & 0.70 & 1.74 & 1.11 \\
 &                     & 0.3 & 0.30 & 0.34 & 0.63 & 0.68 & 0.29 & 0.33 & 0.73 & 0.86 \\
 &                     & 0.5 & 0.21 & 0.30 & 0.59 & 0.66 & 0.26 & 0.28 & 0.78 & 0.87 \\
\cmidrule(lr){2-11}
 & \multirow{3}{*}{10} & 0.1 & 0.53 & 0.57 & 2.45 & 2.99 & 0.63 & 0.75 & 1.56 & 2.23 \\
 &                     & 0.3 & 0.25 & 0.35 & 0.79 & 0.83 & 0.21 & 0.39 & 0.69 & 0.77 \\
 &                     & 0.5 & 0.12 & 0.16 & 0.61 & 0.72 & 0.19 & 0.19 & 0.71 & 0.74 \\
\cmidrule(lr){2-11}
 & \multirow{3}{*}{50} & 0.1 & 0.54 & 0.51 & 0.64 & 0.64 & 0.74 & 0.71 & 0.72 & 0.73 \\
 &                     & 0.3 & 0.25 & 0.30 & 0.61 & 0.59 & 0.23 & 0.28 & 0.70 & 0.69 \\
 &                     & 0.5 & 0.09 & 0.09 & 0.52 & 0.54 & 0.18 & 0.20 & 0.63 & 0.66 \\
\midrule
\multirow{9}{*}{\rotatebox[origin=c]{90}{\textbf{Ang.}}}
 & \multirow{3}{*}{2}  & 0.1 & 0.72 & 0.76 & 0.73 & 0.80 & 0.69 & 0.65 & 0.49 & 0.52 \\
 &                     & 0.3 & 0.24 & 0.45 & 0.24 & 0.50 & 0.23 & 0.35 & 0.27 & 0.39 \\
 &                     & 0.5 & 0.11 & 0.16 & 0.11 & 0.22 & 0.22 & 0.21 & 0.27 & 0.27 \\
\cmidrule(lr){2-11}
 & \multirow{3}{*}{5}  & 0.1 & 0.83 & 0.82 & 0.71 & 0.81 & 0.74 & 0.69 & 0.66 & 0.64 \\
 &                     & 0.3 & 0.40 & 0.62 & 0.31 & 0.54 & 0.18 & 0.39 & 0.19 & 0.31 \\
 &                     & 0.5 & 0.11 & 0.21 & 0.11 & 0.20 & 0.18 & 0.18 & 0.20 & 0.20 \\
\cmidrule(lr){2-11}
 & \multirow{3}{*}{10} & 0.1 & 0.78 & 0.75 & 0.68 & 0.63 & 0.80 & 0.75 & 0.78 & 0.76 \\
 &                     & 0.3 & 0.38 & 0.60 & 0.19 & 0.41 & 0.20 & 0.47 & 0.20 & 0.16 \\
 &                     & 0.5 & 0.10 & 0.20 & 0.18 & 0.13 & 0.19 & 0.19 & 0.20 & 0.20 \\
\bottomrule
\end{tabular}
\end{table}

\begin{table}[!t]
\centering
\caption{\textbf{QUID Attack Success Rate (\%)} across encodings, layers, and poison ratios under noiseless (NL) and depolarized noise (DN, $p=0.01$) conditions, with and without Q-Detection (w/, w/o).}
\label{tab:quid_asr_combined}
\footnotesize
\setlength{\tabcolsep}{2.5pt}
\renewcommand{\arraystretch}{1.1}
\begin{tabular}{@{}c|c|c|cc|cc|cc|cc@{}}
\toprule
\multirow{3}{*}{\textbf{Enc.}} & \multirow{3}{*}{\textbf{L}} & \multirow{3}{*}{\shortstack{\textbf{Poison}\\\textbf{Ratio}}} & \multicolumn{4}{c|}{\textbf{AZ-Class (ASR)}} & \multicolumn{4}{c}{\textbf{MNIST (ASR)}} \\ \cline{4-11}
& & & \multicolumn{2}{c|}{\textbf{NL}} & \multicolumn{2}{c|}{\textbf{DN}} & \multicolumn{2}{c|}{\textbf{NL}} & \multicolumn{2}{c}{\textbf{DN}} \\ \cline{4-11}
& & & \textbf{w/o} & \textbf{w/} & \textbf{w/o} & \textbf{w/} & \textbf{w/o} & \textbf{w/} & \textbf{w/o} & \textbf{w/} \\
\midrule
\multirow{9}{*}{\rotatebox[origin=c]{90}{\textbf{Amp.}}}
 & \multirow{3}{*}{2}  & 0.1 & 70.50 & 58.12 & 76.52 & 58.14 & 59.62 & 23.38 & 73.71 & 40.43 \\
 &                     & 0.3 & 88.12 & 85.58 & 81.05 & 61.03 & 75.79 & 64.16 & 86.10 & 82.81 \\
 &                     & 0.5 & 91.49 & 88.94 & 86.22 & 67.06 & 82.02 & 77.66 & 92.72 & 88.92 \\
\cmidrule(lr){2-11}
 & \multirow{3}{*}{10} & 0.1 & 56.54 & 42.90 & 81.12 & 73.98 & 37.90 & 12.00 & 45.57 & 14.71 \\
 &                     & 0.3 & 87.03 & 79.05 & 93.40 & 84.74 & 85.13 & 56.03 & 87.67 & 86.00 \\
 &                     & 0.5 & 93.15 & 91.53 & 96.43 & 87.02 & 87.08 & 86.73 & 86.43 & 83.69 \\
\cmidrule(lr){2-11}
 & \multirow{3}{*}{50} & 0.1 & 56.85 & 57.72 & 92.98 & 93.98 & 26.64 & 16.50 & 86.43 & 86.43 \\
 &                     & 0.3 & 84.71 & 79.16 & 96.06 & 94.22 & 84.83 & 74.81 & 87.67 & 87.67 \\
 &                     & 0.5 & 93.12 & 93.23 & 98.91 & 97.01 & 87.13 & 87.06 & 89.02 & 88.17 \\
\midrule
\multirow{9}{*}{\rotatebox[origin=c]{90}{\textbf{Ang.}}}
 & \multirow{3}{*}{2}  & 0.1 & 16.25 & 9.79  & 20.93 & 13.08 & 3.00  & 2.76  & 3.90  & 3.86  \\
 &                     & 0.3 & 76.87 & 42.98 & 78.92 & 49.79 & 52.30 & 38.49 & 52.24 & 42.08 \\
 &                     & 0.5 & 88.83 & 82.04 & 88.80 & 83.70 & 53.64 & 54.72 & 52.09 & 52.22 \\
\cmidrule(lr){2-11}
 & \multirow{3}{*}{5}  & 0.1 & 11.29 & 7.58  & 27.35 & 17.31 & 3.90  & 2.90  & 6.67  & 5.05  \\
 &                     & 0.3 & 68.37 & 33.43 & 78.65 & 48.61 & 54.70 & 8.02  & 60.38 & 12.37 \\
 &                     & 0.5 & 85.26 & 69.48 & 84.94 & 74.24 & 53.99 & 54.93 & 54.94 & 54.53 \\
\cmidrule(lr){2-11}
 & \multirow{3}{*}{10} & 0.1 & 14.51 & 10.12 & 62.80 & 41.06 & 3.52  & 5.62  & 13.76 & 11.48 \\
 &                     & 0.3 & 69.31 & 30.57 & 86.82 & 75.69 & 62.22 & 16.16 & 70.92 & 33.60 \\
 &                     & 0.5 & 87.11 & 78.50 & 88.86 & 89.89 & 48.98 & 48.80 & 48.41 & 48.41 \\
\bottomrule
\end{tabular}
\end{table}

QUID~\cite{Kundu-poisoning-2025} poisons training labels by using the distance between quantum-encoded states in Hilbert space, without access to the victim model’s training or gradients. For each poisoned sample, it computes the Frobenius distance between the sample’s encoded density matrix and class prototype density matrices, then flips the label to the most distant class to maximally disrupt class separation. Because QUID relies only on the encoding circuit, it is less tied to training-specific behavior than gradient-based poisoning. To scale the attack to the MNIST and AZ-Class datasets across all encodings and circuit depths, we use class prototype density matrices precomputed from the training data instead of per-sample density matrices. This preserves the original threat model, since label reassignment is driven by class-level geometric separation in Hilbert space.

To mitigate QUID, we adopt Q-Detection~\cite{he2025q}, a hybrid quantum-classical defense that formulates poisoning detection as a bi-level QUBO problem. It trains a Quantum Weight-Assigning Network (Q-WAN) to assign sample weights from per-sample losses, separating likely poisoned from clean samples. The method alternates between \textit{adversarial filtering}, which suppresses corrupted samples by maximizing their weighted loss, and \textit{selective learning}, which emphasizes clean samples by minimizing their weighted loss. Q-WAN parameters are updated from differences between free and guided spin states. For scalability, we use simulated annealing instead of quantum annealing, following the original framework, with $\eta=0.05$, $\alpha=1.0$, $\beta \in [0.1, 2.0]$, and 50 annealing sweeps. We evaluate poison ratios $\varepsilon \in \{0.1, 0.3, 0.5\}$, comparing QUID with and without Q-Detection under both noiseless and depolarizing noise ($p=0.01$).

\paragraphS{Performance under Noiseless Setting}
Table~\ref{tab:quid_accuracy_ratio} and and  Table \ref{tab:quid_asr_combined}report relative accuracy and ASR respectively. Figure \ref{fig:quid-asr} shows the ASR values for QUID attack. Amplitude encoding is highly vulnerable, with degradation increasing with depth. At $\varepsilon=0.5$, relative accuracy falls from about 0.2 to below 0.1, while ASR exceeds 90\% on AZ-Class, indicating near-complete attack success. This is driven by strong class separation in Hilbert space, which helps QUID assign maximally harmful labels, and by weaker clean baselines that increase sensitivity to perturbation. At $\varepsilon=0.1$, relative accuracy remains around 0.5 and ASR is lower, showing a threshold effect consistent with~\cite{Kundu-poisoning-2025}. Q-Detection provides only modest gains, since loss saturation among poisoned samples limits discrimination.

Angle encoding is more robust rhan amplitude encoding at low poison ratios: at $\varepsilon=0.1$, relative accuracy stays around 0.7- 0.8 and ASR remains low. At $\varepsilon=0.5$, accuracy drops to about 0.1- 0.2 and ASR rises to about 85\%, but without a clear depth trend. Q-Detection is more effective here: at $\varepsilon=0.3$, relative accuracy improves substantially and ASR drops sharply. These gains arise from a more graded loss distribution, which helps Q-WAN separate poisoned from clean samples. Gains are smaller on MNIST due to higher intrinsic loss variance. Overall, stronger Hilbert-space separation in amplitude encoding increases both attack success and detection difficulty, while angle encoding limits attack effectiveness and supports more effective mitigation.

\paragraphS{Performance under Depolarized Noise}
Table~\ref{tab:quid_accuracy_ratio} and and  Table \ref{tab:quid_asr_combined}report relative accuracy and ASR respectively. Figure \ref{fig:quid-asr} shows the ASR values for QUID attack. Under noise, QUID is less effective with amplitude encoding. At $\varepsilon=0.5$, relative accuracy rises to about 0.6--0.8 (from about 0.1- 0.3 without noise) in most cases, while ASR drops slightly, showing that noise weakens the poisoning effect. This is because noise disrupts the encoded state patterns that QUID relies on for relabeling~\cite{Mottonen2004, Mottonen2005StateTransformation, Kundu-poisoning-2025}. At $\varepsilon=0.1$, ratios above 1 on AZ-Class reflect collapse of the clean baseline, not true robustness. At greater depth, noise dominates, reducing accuracy and inflating ASR through near-random predictions. Q-Detection remains ineffective because noise makes poisoned and clean samples harder to separate.

Angle encoding is much less affected by noise. At $\varepsilon = 0.5$, relative accuracy and ASR remain close to the noiseless case, showing that the attack largely persists. At $\varepsilon = 0.1$, the effect of noise on relative accuracy is mixed and configuration-dependent. On AZ-Class, layer 2 shows a marginal improvement under noise, while layers 5 and 10 show moderate declines. On MNIST, the drop is more pronounced at layer 2 but minimal at layer 10. This variability reflects the competing effects of noise at low poison ratios: when the poisoning signal is already weak, noise can partially disrupt QUID's geometric label assignments--  providing a marginal natural defense in some configurations-- but also independently degrades model accuracy, which is reflected more prominently in the relative accuracy ratio when absolute accuracy is already modest. The net effect depends on which of these two forces dominates at a given depth and dataset, producing the inconsistent pattern observed across configurations. Q-Detection remains useful under noisy angle encoding at moderate poison ratios, with meaningful ASR reductions at $\varepsilon = 0.3$ across all depths, though gains shrink at deeper layers where noise-induced loss ambiguity makes sample weighting less reliable. Overall, noise weakens QUID on amplitude encoding but also limits detection, while angle encoding preserves both attack strength and partial detectability.

\paragraphS{Statistical Analysis} 
Standard errors remain below $\pm$4 percentage points for ASR across all configurations, with 95\% confidence intervals computed via the $t$-distribution. Amplitude encoding yields significantly higher ASR than angle encoding across all poison ratios on AZ-Class under noiseless conditions (Welch's $t$-test, $p=0.006$ at $\varepsilon=0.1$, $p=0.02$ at $\varepsilon=0.3$, $p=0.017$ at $\varepsilon=0.5$), confirming that stronger Hilbert-space class separation under amplitude encoding meaningfully increases attack success.

Under depolarizing noise, the noiseless-to-noisy ASR difference is significant for amplitude encoding at $\varepsilon=0.3$ and $\varepsilon=0.5$ at layers 2 and 50 ($p\leq0.035$), confirming genuine QUID weakening, while the difference at layer 2, $\varepsilon=0.1$ is not significant ($p=0.092$), consistent with the near-zero AZ-Class baseline producing numerically unstable ratios rather than a true signal. At layer 50, all noise-induced ASR changes are significant ($p\leq0.002$), confirming that cumulative noise overwhelms the amplitude encoding's Hilbert-space structure. 

For angle encoding at $\varepsilon=0.5$, the noiseless-to-noisy ASR difference is non-significant at layers 5 and 10 ($p=0.543$ and $p=0.407$ respectively), confirming that QUID's effectiveness is preserved under noise for angle encoding, as claimed. 

Q-Detection significantly reduces ASR for angle encoding at $\varepsilon=0.3$ across all depths (paired $t$-test, $p\leq0.002$), but provides no significant improvement at $\varepsilon=0.5$ at layers 2 and 10 ($p=0.243$ and $p=0.171$), consistent with the high poison volume overwhelming Q-WAN's reweighting capacity.
For amplitude encoding, Q-Detection improvements are 
significant at lower poison ratios but become non-significant at layer 50 for $\varepsilon=0.1$ and $\varepsilon=0.5$ ($p=0.663$ and $p=0.638$), confirming negligible Q-Detection benefit when the noise-degraded baseline leaves no meaningful loss distribution to exploit.

\subsection{Gray-box $\rightarrow$ HZ Backdoor Attack~\cite{huang2023backdoor}}
\label{sec:huang-eval}

\begin{figure*}[!ht]
    \centering

    \begin{subfigure}[b]{0.24\textwidth}
        \centering
        \includegraphics[width=\textwidth]{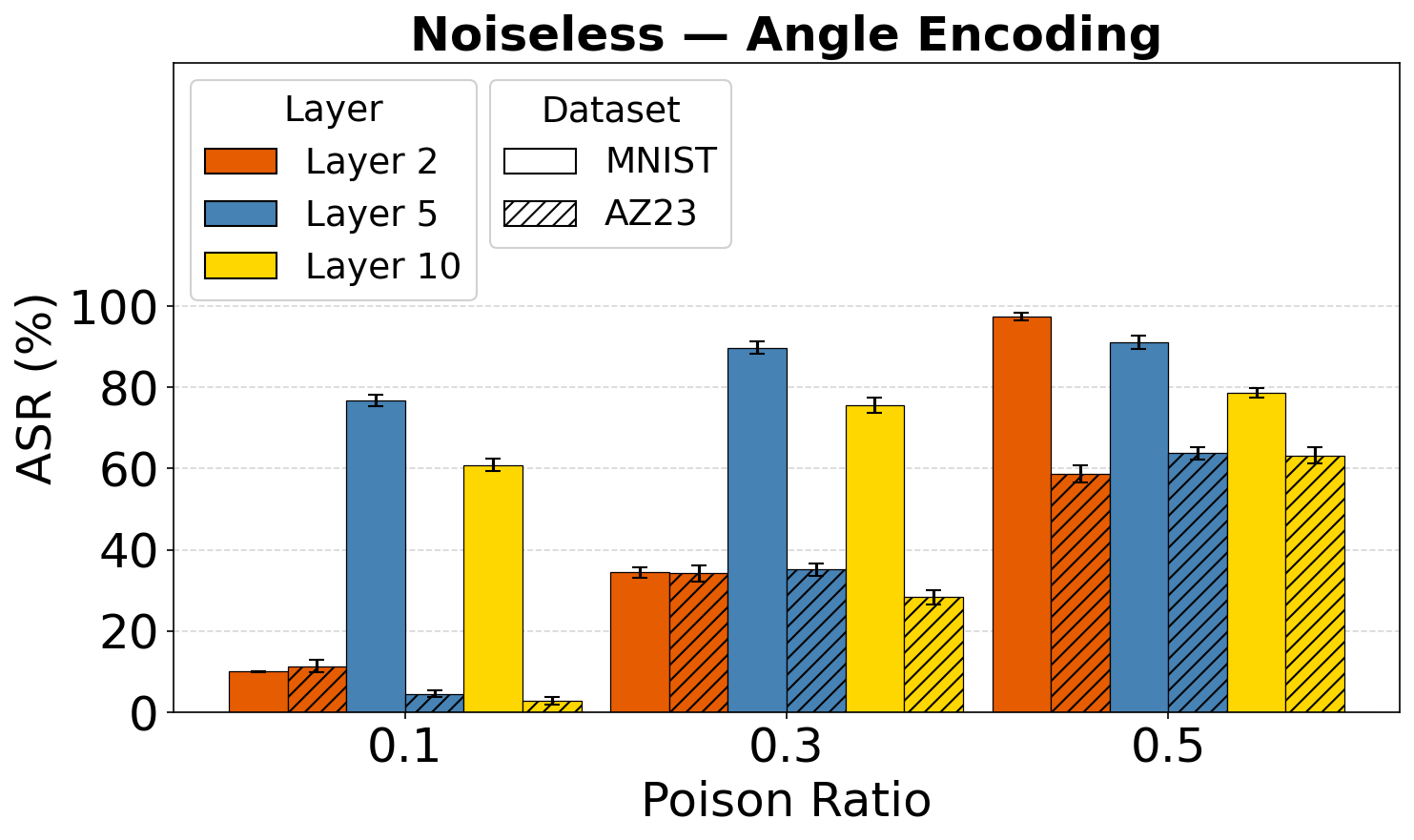}
    \end{subfigure}
    \hfill
    \begin{subfigure}[b]{0.24\textwidth}
        \centering
        \includegraphics[width=\textwidth]{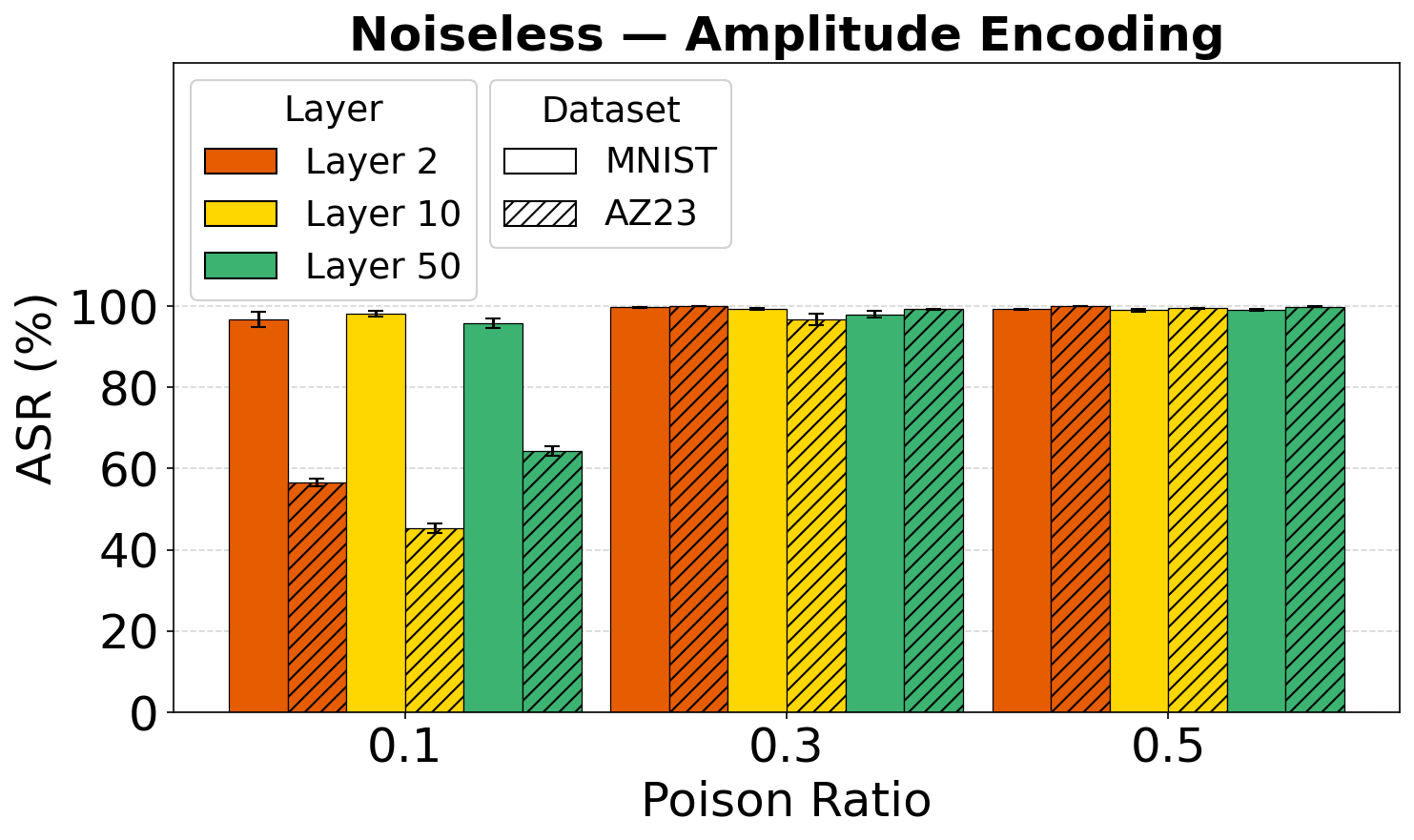}
    \end{subfigure}
    \hfill
    \begin{subfigure}[b]{0.24\textwidth}
        \centering
    
        \includegraphics[width=\textwidth]{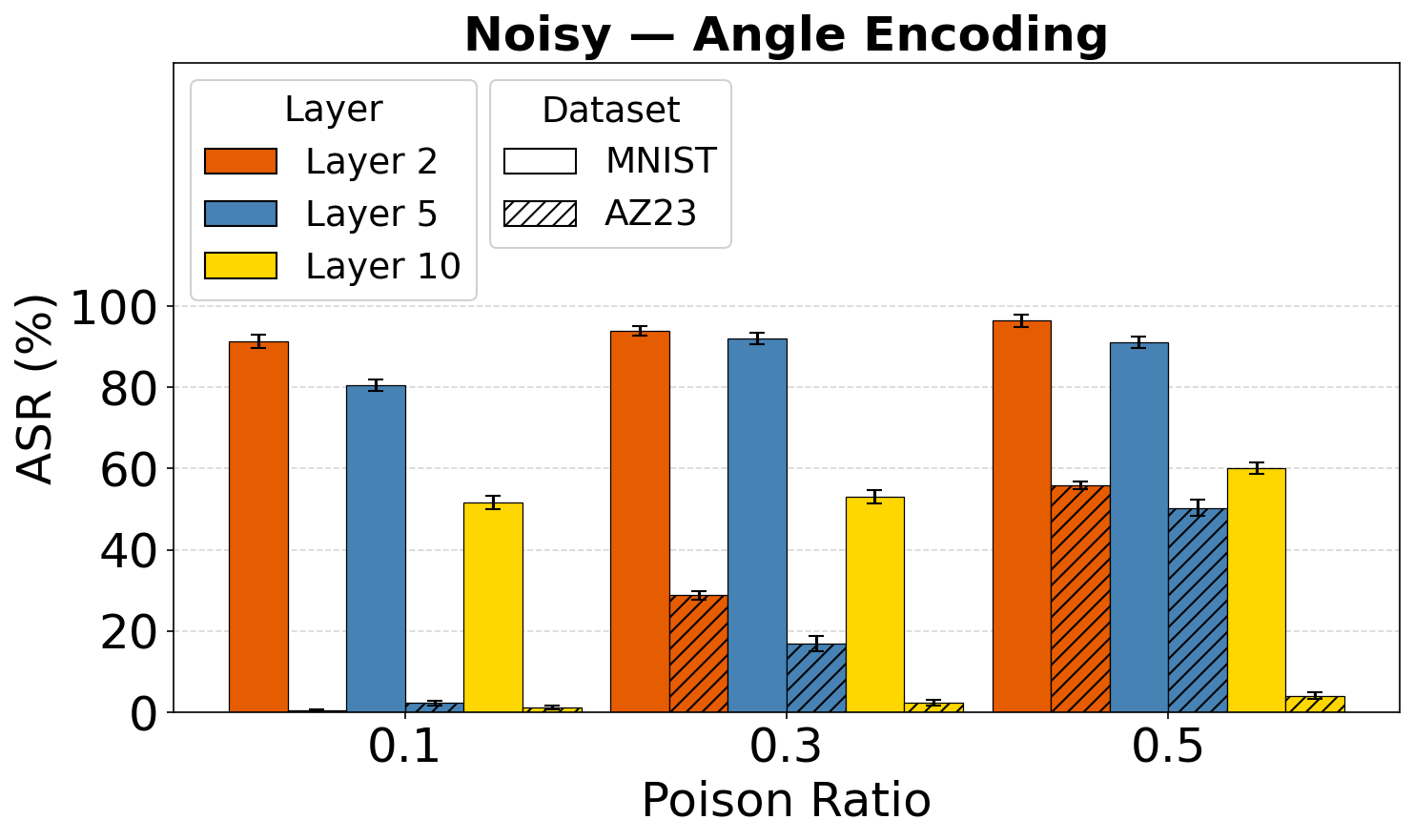}
    \end{subfigure}
    \hfill
    \begin{subfigure}[b]{0.24\textwidth}
        \centering
        \includegraphics[width=\textwidth]{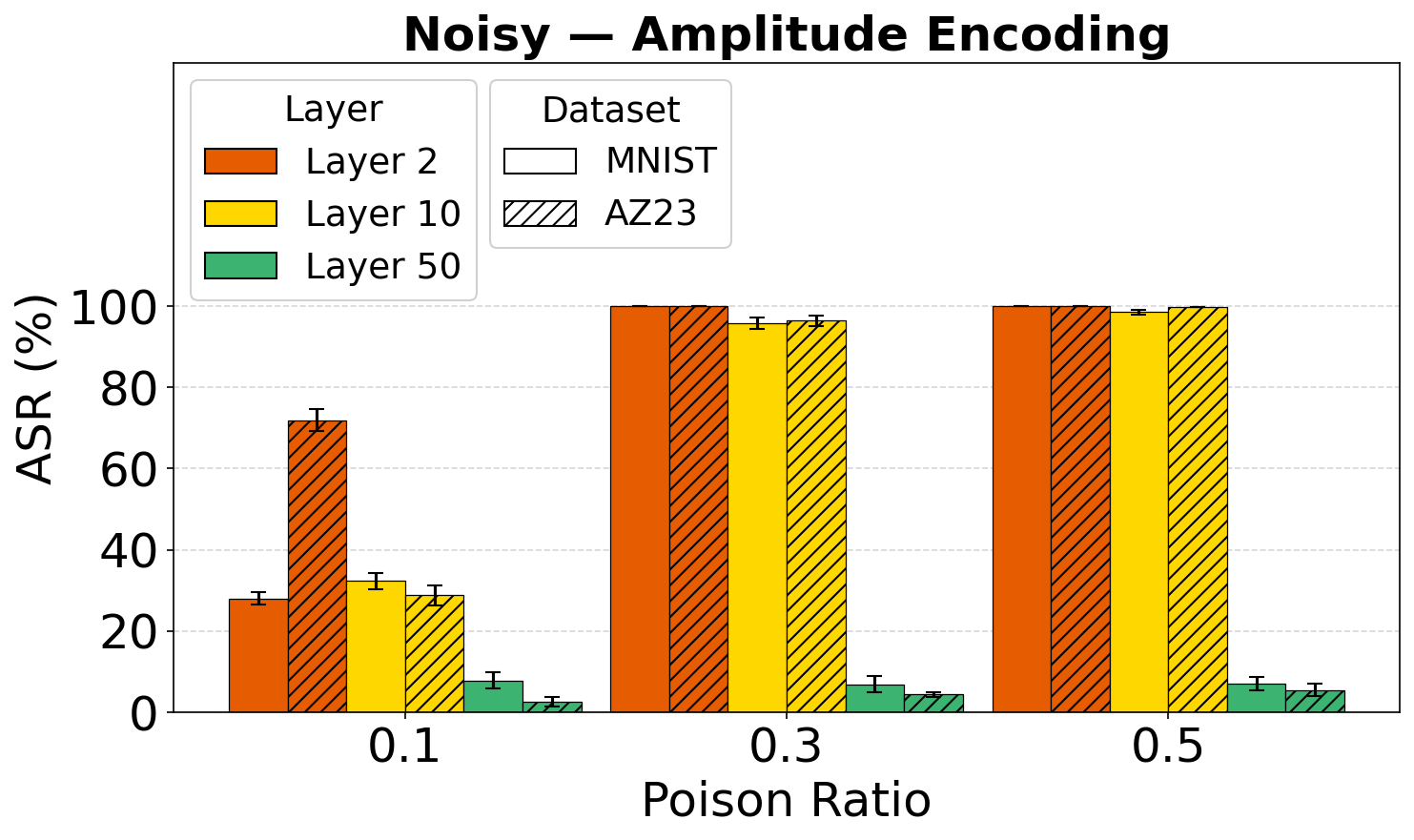}
    \end{subfigure}

    \caption{Attack Success Rate (ASR) of the Huang backdoor attack with amplitude and angle encodings under noiseless and noisy settings with varying layer. 
    }
    \label{fig:huang-asr}
\end{figure*}

The Huang and Zhang (HZ) backdoor~\cite{huang2023backdoor} is a gray-box clean-label poisoning attack that assumes access only to the training data. The attacker trains a same-architecture proxy model, generates a universal trigger $\delta_b$ ($\|\delta_b\|_\infty \leq \varepsilon$) using fuzzy admix and Q-FGSM, and poisons a fraction of target-class samples without changing their labels. We evaluate poison ratios $\{0.1, 0.3, 0.5\}$ for amplitude encoding (2, 10, 50 layers) and angle encoding (2, 5, 10 layers) under both noiseless and depolarizing noise ($p=0.01$). Results are shown in Table~\ref{tab:huang_combined} and Figure~\ref{fig:huang-asr}.

\begin{tcolorbox}[
colback=blue!5,
colframe=blue!40!black,
colbacktitle=blue!15!white,
coltitle=black,
fonttitle=\bfseries,
title filled,
boxrule=0.8pt,
arc=2mm,
title={\em Key Insights: Backdoor Attack~\cite{huang2023backdoor}}
]
\begin{itemize}
    \item The attack remains highly stealthy across encodings, depths, and poison ratios.
    \item Under noiseless conditions, amplitude encoding is markedly more vulnerable because its single-shot embedding gives the optimized trigger a stable entry point.
    \item Angle encoding remains comparatively resilient across depths, with ASR under noise close to noiseless levels.
    \item Unlike QUID, whose effectiveness depends on state-space structure degraded by noise, the Huang trigger is embedded in the trained weights and thus persists as long as the model remains functional.
\end{itemize}
\end{tcolorbox}

\paragraphB{Performance under Noiseless Setting} The attack remains stealthy across all configurations: when the trigger is off, the relative clean data accuracy (CDA) stays between 0.9 and 1.18 (Table~\ref{tab:huang_combined}), indicating that the backdoor causes no meaningful degradation in clean accuracy regardless of encoding type, circuit depth, or poison ratio.

Amplitude encoding is highly vulnerable when the trigger is on. Relative accuracy stays near 1.0 while ASR reaches or approaches 100\% at ratios 0.3 and 0.5 across all depths on both datasets. This depth-independence holds because AmplitudeEmbedding is applied only once before the variational layers, giving the Q-FGSM trigger a fixed entry point not diluted by per-layer re-uploading. At ratio 0.1, MNIST ASR is already high, while AZ-Class is more moderate, reflecting the greater difficulty of redirecting predictions to a single target across 23 classes. A slight depth sensitivity appears on AZ-Class at ratio 0.1, but it vanishes at higher poison ratios.

Angle encoding exhibits comparatively lower vulnerability, with ASR substantially below that of amplitude encoding across most configurations. On AZ-Class, ASR at ratio 0.5 is moderate and roughly depth-stable, since the clean baseline improves only marginally with depth. On MNIST, where the baseline improves more substantially, ASR at ratio 0.5 declines with depth, consistent with a stronger clean-data signal counteracting the trigger. At ratio 0.1, MNIST ASR is anomalously higher at layers~5 and~10 (76.77\%, 60.88\%) than at layer~2 (10.00\%), an inversion not explained by depth or clean accuracy, which we report as a dataset-architecture interaction requiring further investigation.

\paragraphB{Performance under Depolarized Noise Setting} Under noise, some relative CDA ratios for amplitude encoding appear inflated (up to 4.61 on AZ-Class), not due to improvement but because the clean baseline collapses to near-random accuracy (Table~\ref{tab:baseline-qmlp}). The backdoored model retains higher accuracy, though the cause requires further study.

At poison ratios 0.3 and 0.5, the backdoor remains effective at layers~2 and~10, with ASR above 95\%. At ratio 0.1, behavior differs: ASR increases on AZ-Class but drops sharply on MNIST, indicating that noise can either amplify or weaken the trigger depending on its initial strength. At layer~50, both accuracy and ASR collapse to near-random levels across all ratios, as noise overwhelms the circuit, effectively disabling the attack. Angle encoding remains stable under noise. Relative CDA stays near 1.0, and ASR closely follows noiseless trends, except on AZ-Class at layer~10, where ASR drops to near-random levels at high poison ratio, likely due to difficulty maintaining targeted predictions under noise.

Overall, unlike QUID, which relies on state patterns disrupted by noise, the backdoor attack by Huang et al. \cite{huang2023backdoor} is embedded in model weights and persists unless noise degrades the model itself, as seen at deeper layers.

\begin{table}[!t]
\centering
\caption{HZ Backdoor Attack: Relative CDA and ASR (\%) under noiseless (NL) and depolarizing noise (DN, $p=0.01$) conditions.}
\label{tab:huang_combined}
\footnotesize
\setlength{\tabcolsep}{2.5pt}
\renewcommand{\arraystretch}{1.1}
\begin{tabular}{@{}c|c|c|cc|cc|cc|cc@{}}
\toprule
\multirow{3}{*}{\rotatebox[origin=c]{90}{\textbf{Enc.}}} & \multirow{3}{*}{\textbf{L}} & \multirow{3}{*}{\shortstack{\textbf{Poison}\\\textbf{Ratio}}} & \multicolumn{4}{c|}{\textbf{AZ-Class}} & \multicolumn{4}{c}{\textbf{MNIST}} \\ \cline{4-11}
& & & \multicolumn{2}{c|}{\textbf{Rel. CDA}} & \multicolumn{2}{c|}{\textbf{ASR}} & \multicolumn{2}{c|}{\textbf{Rel. CDA}} & \multicolumn{2}{c}{\textbf{ASR}} \\ \cline{4-11}
& & & \textbf{NL} & \textbf{DN} & \textbf{NL} & \textbf{DN} & \textbf{NL} & \textbf{DN} & \textbf{NL} & \textbf{DN} \\
\midrule
\multirow{9}{*}{\rotatebox[origin=c]{90}{\textbf{Amp.}}}
 & \multirow{3}{*}{2}  & 0.1 & 1.01 & 4.61 & 56.64  & 71.93  & 0.96 & 2.48 & 96.82  & 28.05  \\
 &                     & 0.3 & 0.98 & 4.59 & 99.94  & 100.00 & 0.92 & 2.45 & 99.73  & 100.00 \\
 &                     & 0.5 & 0.98 & 4.46 & 99.96  & 100.00 & 0.9 & 2.34 & 99.23  & 100.00 \\
\cmidrule(lr){2-11}
 & \multirow{3}{*}{10} & 0.1 & 0.98 & 3.79 & 45.27  & 28.84  & 1.18 & 2.14 & 98.20  & 32.33  \\
 &                     & 0.3 & 0.96 & 3.78 & 96.66  & 96.39  & 1.18 & 2.18 & 99.25  & 95.72  \\
 &                     & 0.5 & 0.96 & 3.68 & 99.51  & 99.75  & 1.17 & 2.18 & 99.01  & 98.55  \\
\cmidrule(lr){2-11}
 & \multirow{3}{*}{50} & 0.1 & 1.01 & 0.57 & 64.34  & 2.58   & 1.0 & 0.42 & 95.79  & 7.88   \\
 &                     & 0.3 & 1.03 & 0.71 & 99.26  & 4.39   & 1.00 & 0.48 & 98.00  & 6.94   \\
 &                     & 0.5 & 0.98 & 0.76 & 99.90  & 5.50   & 0.99 & 0.49 & 99.02  & 7.06   \\
\midrule
\multirow{9}{*}{\rotatebox[origin=c]{90}{\textbf{Ang.}}}
 & \multirow{3}{*}{2}  & 0.1 & 0.98 & 0.98 & 11.40  & 0.48   & 1.15 & 1.17 & 10.00  & 91.40  \\
 &                     & 0.3 & 0.98 & 0.97 & 34.26  & 28.80  & 1.16 & 1.16 & 34.42  & 93.97  \\
 &                     & 0.5 & 0.97 & 0.97 & 58.72  & 55.90  & 1.04 & 1.05 & 97.50  & 96.39  \\
\cmidrule(lr){2-11}
 & \multirow{3}{*}{5}  & 0.1 & 1.01 & 1.02 & 4.55   & 2.29   & 1.02 & 1.00 & 76.77  & 80.55  \\
 &                     & 0.3 & 1.02 & 1.03 & 35.20  & 16.88  & 0.98 & 0.98 & 89.81  & 92.03  \\
 &                     & 0.5 & 1.01 & 1.02 & 63.80  & 50.28  & 0.99 & 1.01 & 91.07  & 91.16  \\
\cmidrule(lr){2-11}
 & \multirow{3}{*}{10} & 0.1 & 0.99 & 0.99 & 2.83   & 1.26   & 1.07 & 1.02 & 60.88  & 51.64  \\
 &                     & 0.3 & 0.99 & 0.99 & 28.32  & 2.38   & 1.08 & 1.02 & 75.58  & 53.08  \\
 &                     & 0.5 & 0.96 & 1.0 & 63.25  & 4.10   & 1.03 & 0.82 & 78.64  & 60.10  \\
\bottomrule
\end{tabular}
\end{table}

\paragraphB{Statistical Analysis} 
The attack remains stealthy: in 22 of 36 configurations, CDA does not differ significantly from the clean baseline (paired $t$-test, $p{>}0.05$). Amplitude encoding yields significantly higher ASR than angle encoding (Welch's $t$-test, $p{=}0.0001{<}0.05$ at $\varepsilon{=}0.1$), except at $\varepsilon{=}0.5$ on MNIST where both saturate ($p{=}0.2734{>}0.05$). 

Under noise, inflated relative clean data accuracy ratios for amplitude encoding at layers~2 and~10 reflect clean baseline collapse rather than improvement. At layer~50, noise collapses both CDA and ASR to near-random levels ($p{=}0.0004{<}0.05$), disabling the attack entirely. 

For angle encoding, ASR remains stable under noise at shallow depths ($p{=}0.2504{>}0.05$ at 2 layers) but degrades significantly at deeper circuits ($p{=}0.0005{<}0.05$ at 10 layers on AZ-Class), where noise scatters predictions away from the target class despite the model remaining functional. The divergent behavior across datasets at low poison ratios- ASR rising on MNIST but collapsing on AZ-Class under identical noise- is consistent with output dimensionality effects: in 10-class MNIST, noise-diffused predictions concentrate on fewer classes, benefiting the target, whereas in 23-class AZ-Class, predictions diffuse too thinly for any single target to accumulate probability mass.

\subsection{White-box $\rightarrow$ Qtrojan Attack}
\label{sec:qtrojan-eval}

\begin{tcolorbox}[
colback=blue!5,
colframe=blue!40!black,
colbacktitle=blue!15!white,
coltitle=black,
fonttitle=\bfseries,
title filled,
boxrule=0.8pt,
arc=2mm,
title={\em Key Insights: Qtrojan}
]
\begin{itemize}
    \item QTrojan preserves CDA across all depths and datasets, making it a stealthy attack.
    \item ASR is consistently low and highly variable across all configurations, contrasting sharply with the 100\% ASR reported in the original paper~\cite{chu2023qtrojan}, due to the increased number of classes in our evaluation.
    \item Depolarizing noise does not degrade CDA but keeps ASR near chance level across all configurations, acting as a natural defense by further randomizing an already-failing forced quantum state. 
\end{itemize}
\end{tcolorbox}

QTrojan~\cite{chu2023qtrojan} is a circuit-level white-box backdoor that inserts two gate layers around the encoding layer $S_x$. When disabled, the circuit is identical to a clean QMLP and CDA is unaffected. When activated, a pre-encoding layer $\bar{S}_x$ applies $RX(\pi/2)$ to every qubit, neutralizing data encoding, and a post-encoding layer $\tilde{S}_x$ applies $RX(3\pi/2) + RY(\theta_\text{target})$ to force all qubits into a fixed attacker-chosen state, redirecting all predictions to a predefined target class. Unlike QUID~(Section~\ref{sec:quid-eval}), and the Huang backdoor~(Section~\ref{sec:huang-eval}), QTrojan operates purely at the circuit level and requires neither training data nor model retraining. Furthermore, because the backdoor is embedded directly into the circuit structure, it persists even if the victim retrains the model on clean data. We classify QTrojan as white-box because the attacker requires full knowledge of the circuit architecture, encoding scheme, and hardware execution stack to insert and activate the backdoor gates at precise positions within each
variational layer.

\begin{figure}[!t]
\centering
\begin{minipage}{0.45\linewidth}
    \centering
    \scriptsize
    \setlength{\tabcolsep}{3pt}
    \begin{tabular}{@{}p{0.35cm}|c|cc|cc@{}}
    \toprule
    \multirow{2}{*}{\rotatebox[origin=c]{90}{\textbf{Dataset}}} & \multirow{2}{*}{\textbf{L}} & \multicolumn{2}{c|}{\textbf{Rel. CDA}} & \multicolumn{2}{c}{\textbf{ASR (\%)}} \\
    \cmidrule(lr){3-4}\cmidrule(lr){5-6}
    & & \textbf{NL} & \textbf{Noisy} & \textbf{NL} & \textbf{Noisy} \\
    \midrule
    \multirow{3}{*}{\rotatebox[origin=c]{90}{\textbf{AZ-23}}}
     & 2  & 0.98 & 1.00 & 7.37  & 6.85  \\
     & 5  & 1.00 & 1.02 & 3.33  & 1.54  \\
     & 10 & 1.01 & 1.03 & 1.63  & 2.54  \\
    \midrule
    \multirow{3}{*}{\rotatebox[origin=c]{90}{\textbf{MNIST}}}
     & 2  & 1.13 & 1.24 & 17.12 & 20.83 \\
     & 5  & 1.01 & 1.09 & 5.55  & 18.25 \\
     & 10 & 1.03 & 1.16 & 6.50  & 1.56  \\
    \bottomrule
    \end{tabular}
    \captionof{table}{QTrojan relative CDA and ASR under noiseless and noisy conditions.}
    \label{tab:qtrojan_combined}
\end{minipage}
\hfill
\begin{minipage}{0.5\linewidth}
    \centering
    \includegraphics[width=\linewidth]{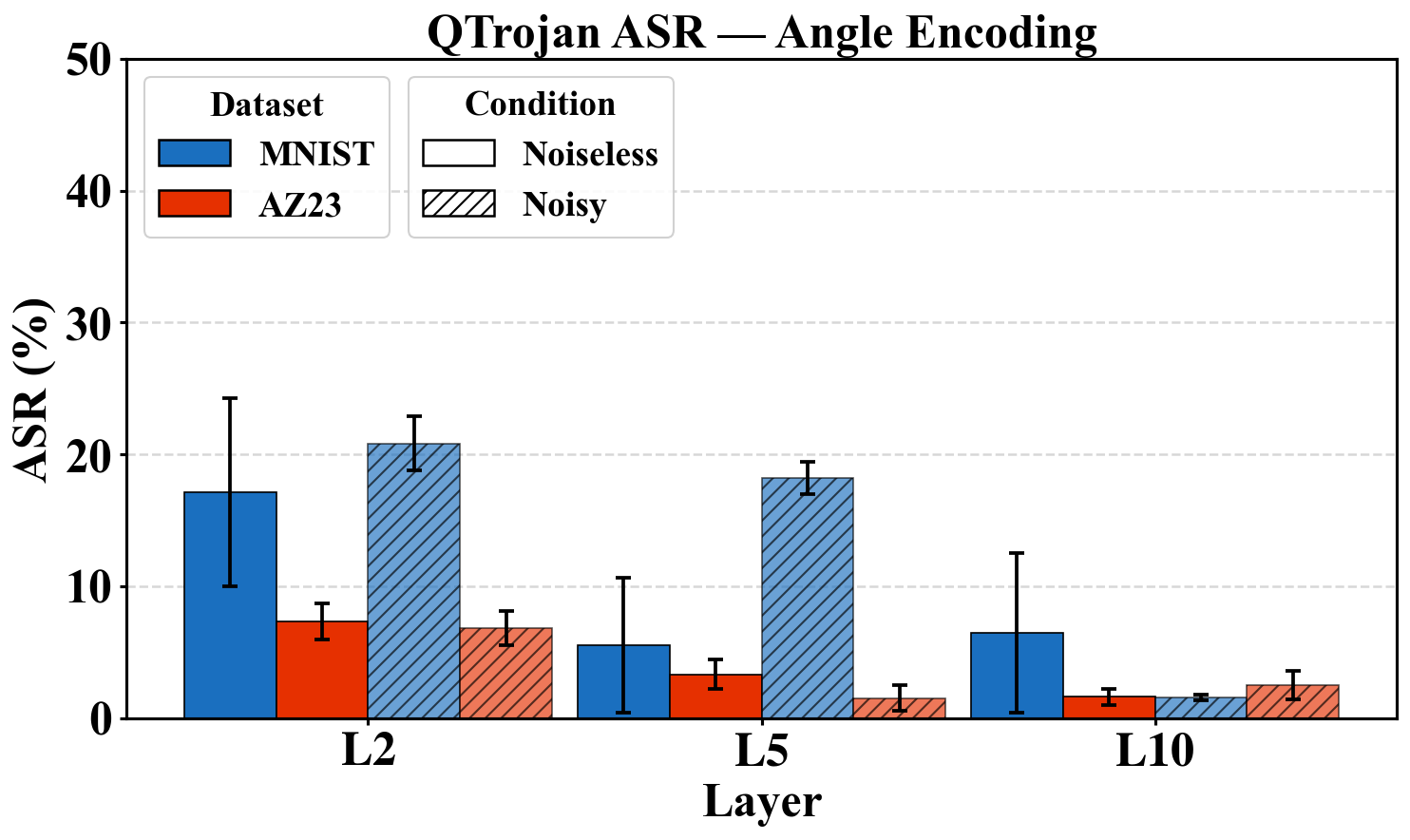}
    \captionof{figure}{ASR of QTrojan across circuit depths under noiseless and noisy conditions using angle encoding.}
    \label{fig:qtrojan}
\end{minipage}
\end{figure}

We evaluate QTrojan with target class~0 and $\theta_\text{target} = \pi/4$ under angle encoding, across layers~2, 5, and~10, on both MNIST and AZ-23. CDA is measured with the backdoor disabled. ASR is the fraction of clean test samples predicted as the target class when the backdoor is active. Results are in Table~\ref{tab:qtrojan_combined}. Figure \ref{fig:qtrojan} shows the ASR(\%) across various layers in both noiseless and noisy setting for both the datasets. As QTrojan operates by manipulating the angle encoding layer $S_x$, it is not applicable to amplitude-encoded models and we therefore restrict evaluation to angle encoding.

\paragraphB{Performance under Noiseless Setting} CDA is preserved across all configurations. Relative CDA ratios range from 0.98 to 1.13, confirming that the inserted gate layers are transparent when the backdoor is disabled, as they reduce to identity under a benign configuration and thus the attack remains stealthy. CDA increases monotonically with depth on both datasets, consistent with the clean baseline trend
(Section~\ref{sec:QMLPBaseline}), confirming that the backdoor circuitry does not interfere with learning.

ASR is substantially lower than the 100\% reported in the original paper~\cite{chu2023qtrojan}, peaking at 17.12\% on MNIST and 7.37\% on AZ-23 at layer~2. For reference, random prediction yields $\sim$10\% ASR on MNIST and $\sim$4.3\% on AZ-23; most observed ASR values fall within or below this range, indicating that QTrojan effectively fails to redirect predictions to the target class in our multi-class setting. 

The attack failure directly follows from task complexity. The original paper\cite{chu2023qtrojan} evaluates 2-class and 4-class MNIST on 16~qubits, where forcing all qubits to $\theta_\text{target}$ sufficiently biases the output toward the target class. In our setting, with 10~classes (MNIST) and 23~classes (AZ-23) on 9~qubits, the fixed quantum state imposed by $\tilde{S}_x$ must overcome competition from many more class-discriminative directions learned by the classical linear layer, which it does not achieve. This reveals that QTrojan's effectiveness degrades significantly with the number of output classes-- a limitation not characterized in the original work.

\begin{table}[!t]

\caption{Relative accuracy of FGSM and PGD attacks for CMLP, QMLP--Angle (Angle), and QMLP--Amplitude (Amp).}
\label{tab:pgd-fgsm_ratio}
\footnotesize
\centering
\setlength{\tabcolsep}{3pt}
\renewcommand{\arraystretch}{1.35}
\begin{tabular}{@{}c|c|ccc|ccc@{}}
\toprule
\multirow{2}{*}{\textbf{Layers}} &
\multirow{2}{*}{\textbf{Attack}} &
\multicolumn{3}{c|}{\textbf{MNIST Ratio}} &
\multicolumn{3}{c}{\textbf{AZ-Class Ratio}} \\
\cmidrule(lr){3-5} \cmidrule(lr){6-8}
 & & \textbf{CMLP} & \textbf{Angle} & \textbf{Amp} & \textbf{CMLP} & \textbf{Angle} & \textbf{Amp} \\
\midrule

\multirow{6}{*}{\textbf{2}}
& FGSM $\epsilon{=}0.01$ &
0.98 & 0.49 & 0.30 &
0.97 & 0.39 & 0.34 \\
& FGSM $\epsilon{=}0.10$ &
0.33 & 0.05 & 0.1 &
0.30 & 0.32 & 0.02 \\
& FGSM $\epsilon{=}0.15$ &
0.17 & 0.01 & 0.04 &
0.13 & 0.28 & 0.01 \\
\cmidrule(lr){2-8}
& PGD $\epsilon{=}0.01$ &
0.98 & 0.48 & 0.27 &
0.97 & 0.39 & 0.33 \\
& PGD $\epsilon{=}0.10$ &
0.09 & 0.04 & 0.03 &
0.10 & 0.31 & 0.01 \\
& PGD $\epsilon{=}0.15$ &
0.02 & 0.01 & 0.00 &
0.02 & 0.26 & 0.00 \\

\midrule
\multirow{6}{*}{\textbf{50}}
& FGSM $\epsilon{=}0.01$ &
0.98 & 0.35 & 0.30 &
0.97 & 0.16 & 0.27 \\
& FGSM $\epsilon{=}0.10$ &
0.33 & 0.07 & 0.00 &
0.30 & 0.15 & 0.000 \\
& FGSM $\epsilon{=}0.15$ &
0.17 & 0.05 & 0.00 &
0.13 & 0.15 & 0.00 \\
\cmidrule(lr){2-8}
& PGD $\epsilon{=}0.01$ &
0.98 & 0.33 & 0.24 &
0.97 & 0.13 & 0.23 \\
& PGD $\epsilon{=}0.10$ &
0.09 & 0.03 & 0.00 &
0.10 & 0.03 & 0.00 \\
& PGD $\epsilon{=}0.15$ &
0.02 & 0.01 & 0.00 &
0.02 & 0.02 & 0.00 \\

\bottomrule
\end{tabular}
\end{table}

\paragraphB{Performance under Depolarized Noise Setting} Absolute CDA values under noise remain close to the noisy baseline across all configurations, with relative ratios ranging from 1.0 to 1.24 (Table~\ref{tab:qtrojan_combined}), confirming that the backdoor circuitry introduces no additional degradation under noise. Since CDA is evaluated with the backdoor disabled, the QTrojan circuit is structurally identical to a clean QMLP at inference, and any noise-induced accuracy decline, is consistent with the general depth-dependent noise sensitivity of angle-encoded circuits established in the baseline (Section~\ref{sec:QMLPBaseline}), and is not attributable to QTrojan.

ASR remains near chance level across all configurations under noisy conditions, as it does in the noiseless setting, confirming that QTrojan fails entirely in our broader multi-class evaluation. The apparent variation across layers and datasets falls within the range of run-to-run variability given the already-low and highly variable noiseless ASR (Table~\ref{tab:qtrojan_combined}), and should not be interpreted as a meaningful behavioral difference. Depolarizing noise acts as a natural defense in this context as well, but not by suppressing a functioning backdoor, rather by further randomizing an already-failing forced quantum state- keeping ASR at chance level. This differs fundamentally from its effect on QUID (Section~\ref{sec:quid-eval}), where noise suppresses a genuinely effective attack by eroding Hilbert-space geometry.

\paragraphB{Statistical Analysis} 
95\% confidence intervals are computed via the $t$-distribution. CDA does not differ significantly from the clean baseline in 11 of 12 configurations (paired $t$-test, $p{>}0.05$), confirming attack stealthiness. ASR differences between noiseless and noisy conditions are non-significant at shallow depths (AZ-23 layer~2: $p{=}0.786$; MNIST layer~2: $p{=}0.641$), indicating trigger persistence under noise. At deeper circuits, noise significantly suppresses ASR (AZ-23 layer~10: $p{=}0.037$; MNIST layer~10: $p{=}0.015$), effectively disabling the attack.

\begin{figure*}[!ht]
    \centering

    \begin{subfigure}[b]{0.24\textwidth}
        \centering
        \includegraphics[width=\textwidth]{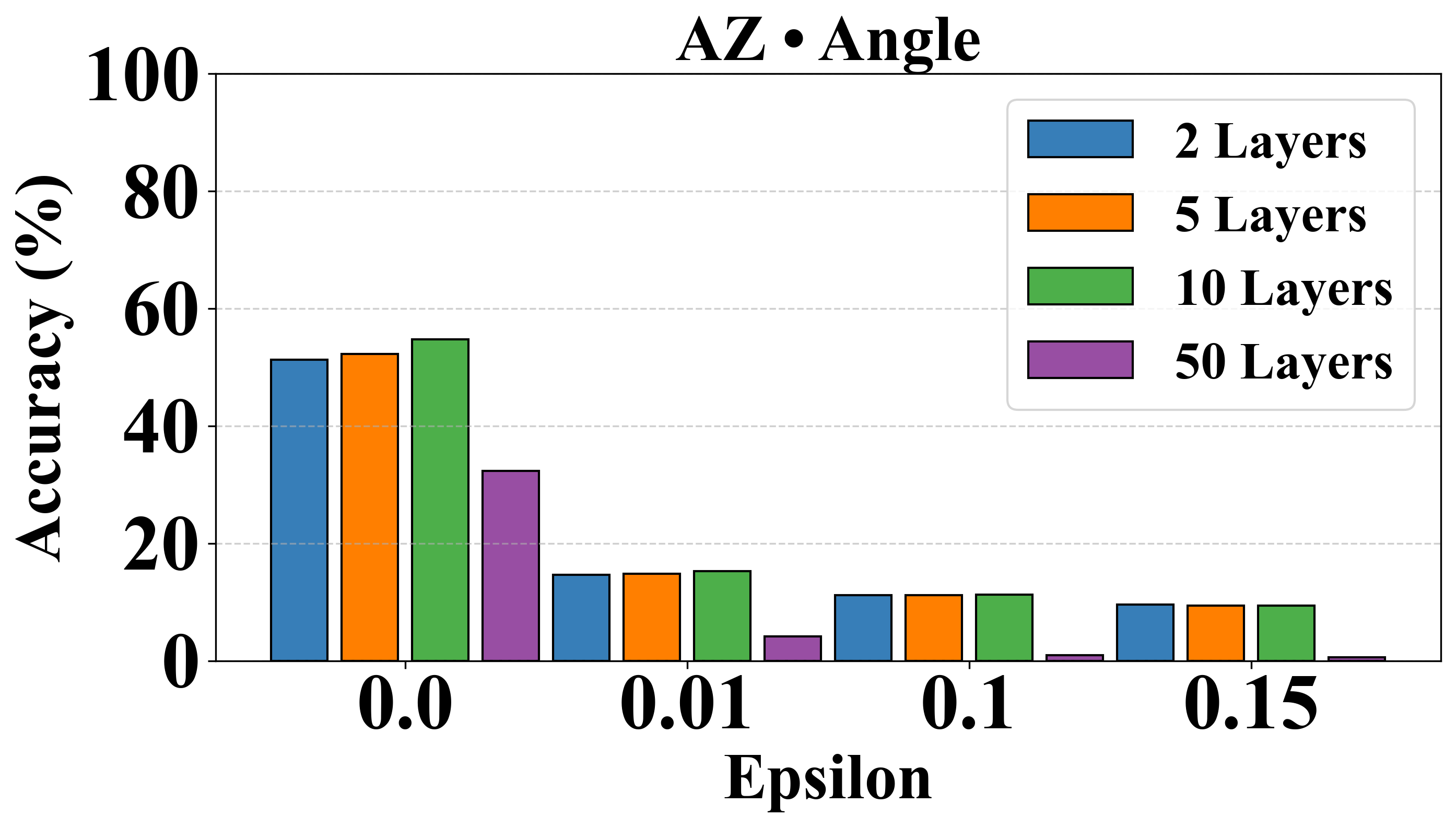}
    \end{subfigure}
    \hfill
    \begin{subfigure}[b]{0.24\textwidth}
        \centering
        \includegraphics[width=\textwidth]{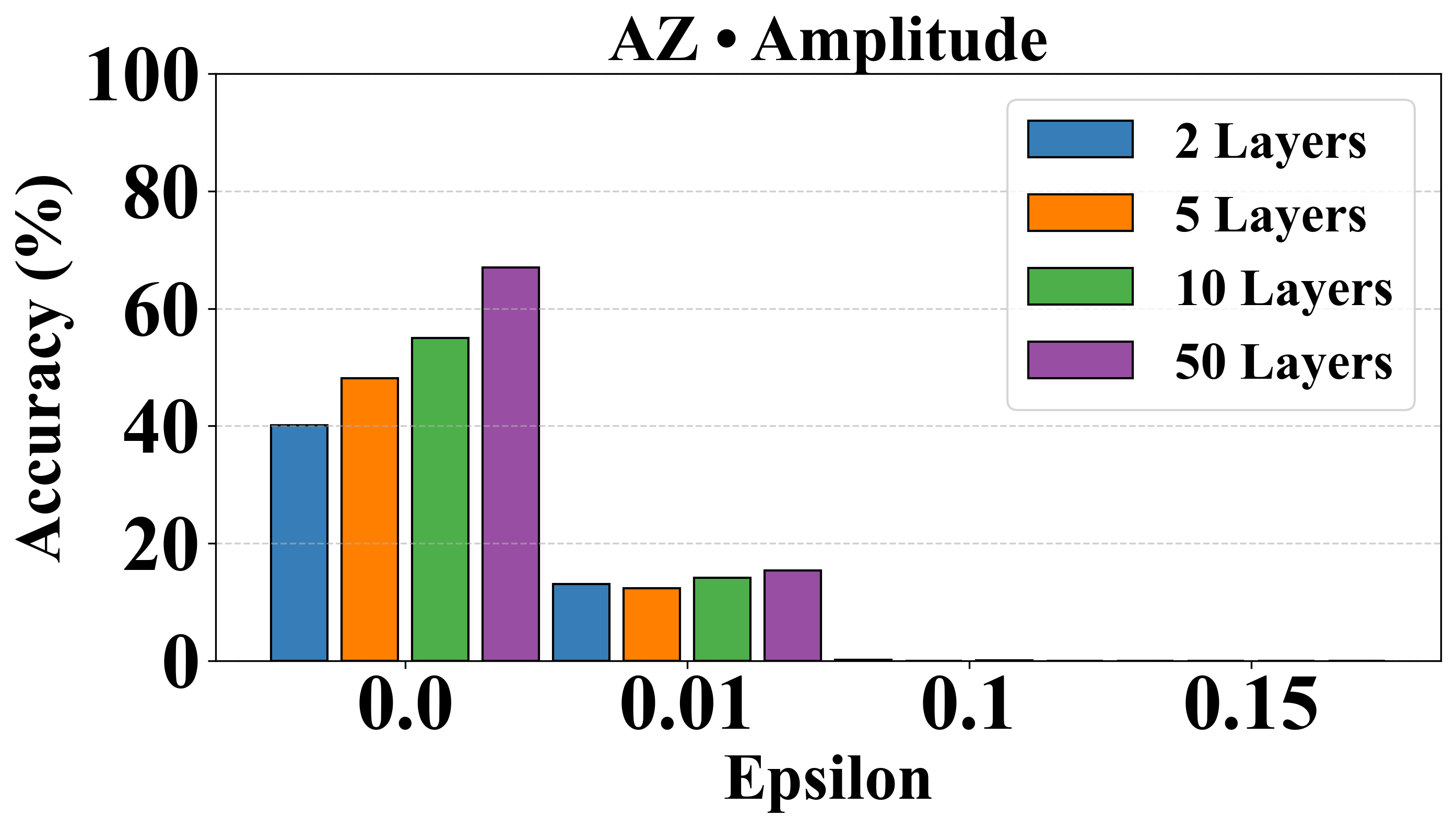}
    \end{subfigure}
    \hfill
    \begin{subfigure}[b]{0.24\textwidth}
        \centering
    
        \includegraphics[width=\textwidth]{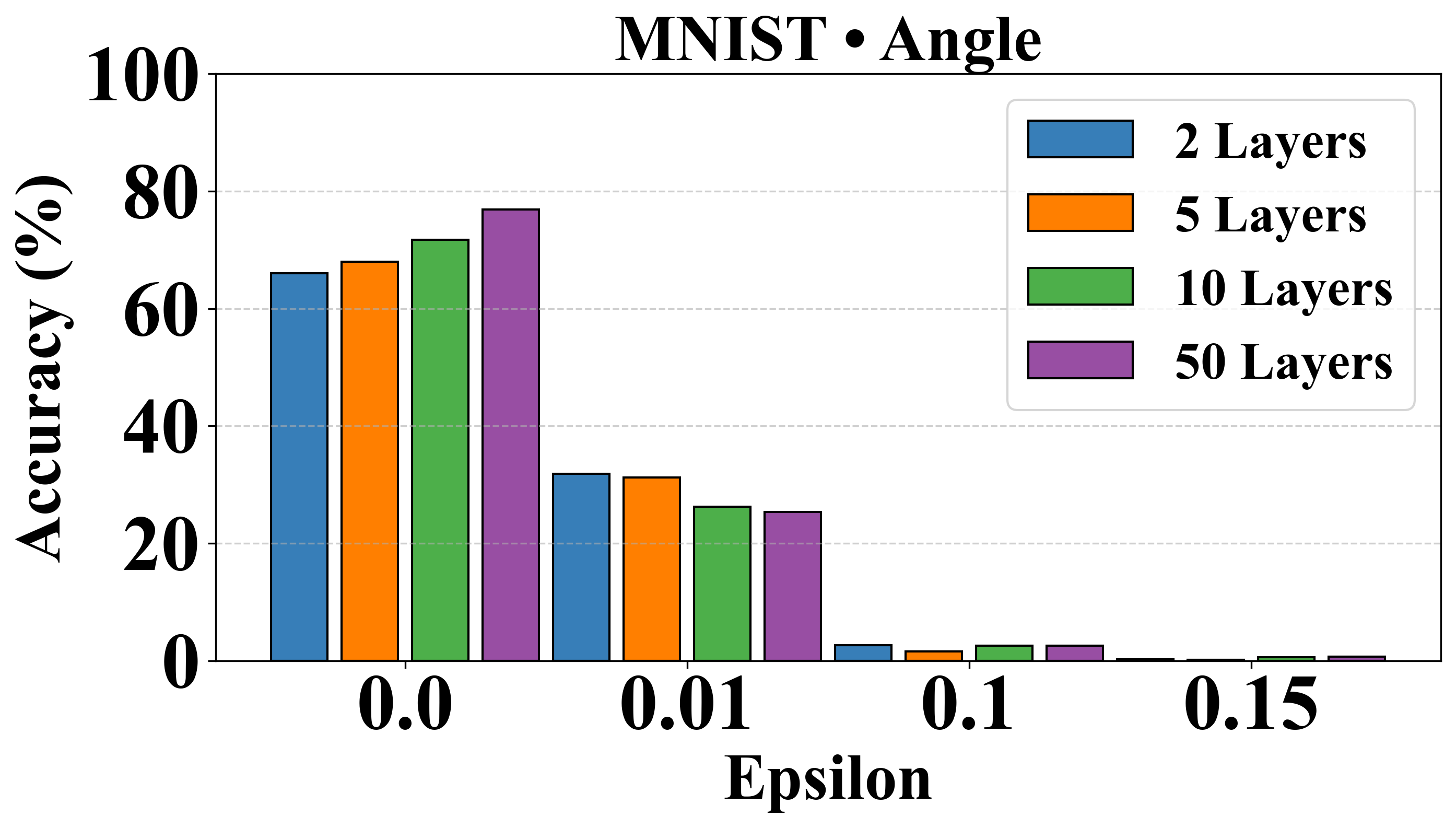}
    \end{subfigure}
    \hfill
    \begin{subfigure}[b]{0.24\textwidth}
        \centering
        \includegraphics[width=\textwidth]{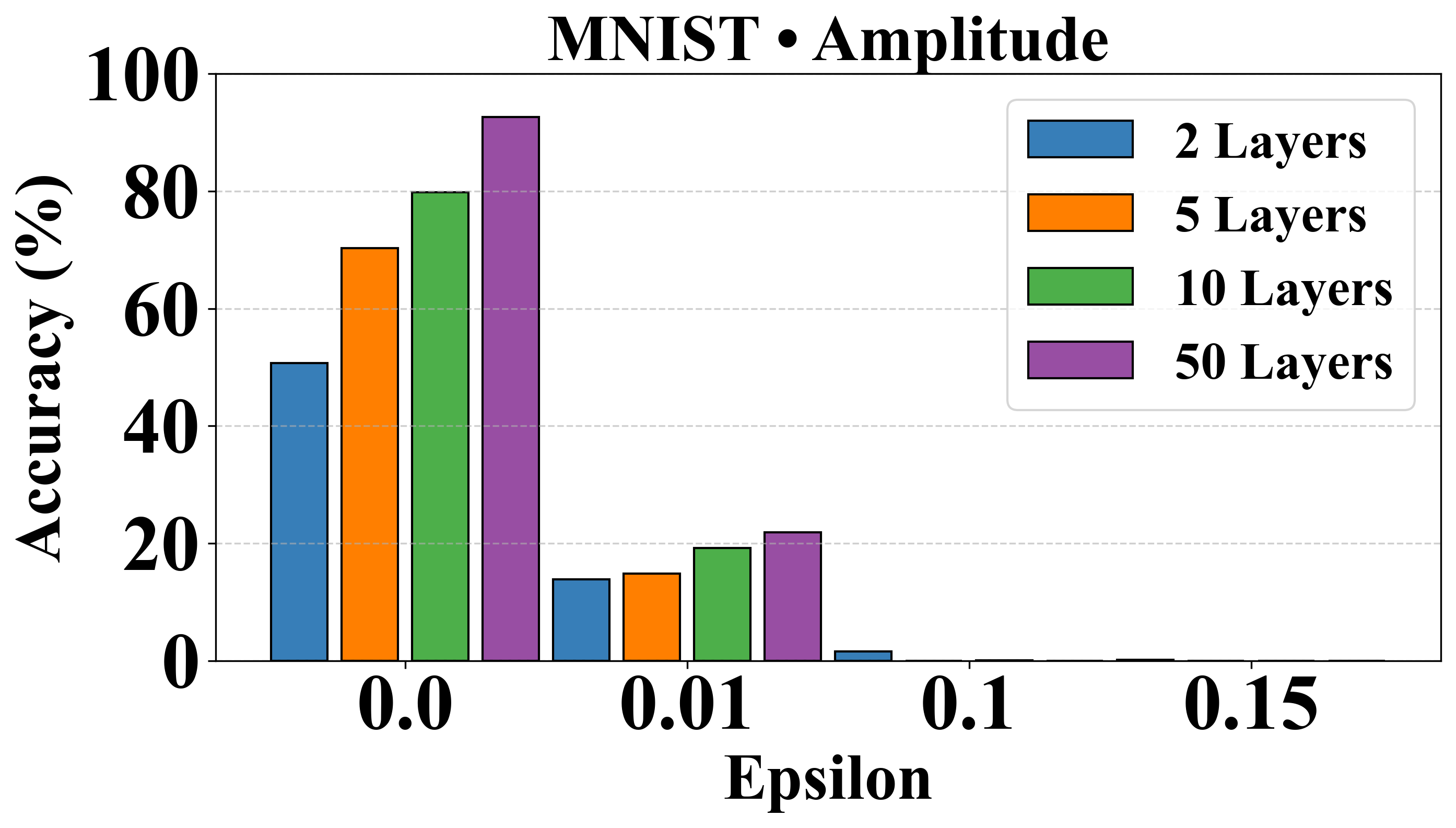}
    \end{subfigure}

    \caption{
    Performance of QMLP with varying perturbation strengths of the PGD attack in a noiseless environment. 
    }
    \label{fig:qmlp_pgd}
\end{figure*}

\subsection{White-box $\rightarrow$ Evasion Attacks}
\label{sec:PGD-FGSM}

\begin{tcolorbox}[
    colback=blue!5,
    colframe=blue!40!black,
    colbacktitle=blue!15!white,
    coltitle=black,
    fonttitle=\bfseries,
    title filled,
    boxrule=0.8pt,
    arc=2mm,
    left=4mm, right=4mm, top=2mm, bottom=2mm,
    title={\em Key Insights: Evasion Attacks}
]
\begin{itemize}
   \item Shallow angle-encoded QMLPs retain moderate robustness;
    deeper circuits collapse due to over-entanglement degrading
    baseline representations before any attack.
    \item Amplitude-encoded models collapse uniformly across all
    depths under small perturbations due to single-shot embedding
    providing a fixed, depth-independent adversarial entry point.
    \item CMLP dominates at $\varepsilon{=}0.01$, but shallow angle-encoded QMLPs outperform it at $\varepsilon{\geq}0.10$ on AZ-Class under both FGSM and PGD, and at layer~50, under FGSM at $\varepsilon{=}0.15$.
\end{itemize}
\end{tcolorbox}

As a representative white-box attack scenario, we evaluate both angle- and amplitude-encoded QMLP models under FGSM~\cite{west2023benchmarking,Lu2020QuantumAdversarial}
and PGD~\cite{west2023benchmarking} attacks. FGSM computes a single-step gradient perturbation scaled by $\varepsilon$, while PGD iterates this process over multiple steps, producing stronger adversarial examples, where $\varepsilon$ controls the maximum allowable perturbation magnitude. We vary $\varepsilon \in
\{0.01, 0.10, 0.15\}$ and circuit depths across MNIST and
AZ-Class under noiseless conditions using angle and amplitude encoding.

\paragraphB{FGSM and PGD Attacks under Noiseless Setting.}
Figure~\ref{fig:qmlp_pgd} and Table~\ref{tab:pgd-fgsm_ratio} summarize results for QMLP (layers 2 and 50) and CMLP.

Robustness degrades sharply with depth for angle encoded models. On AZ-Class, the 2-layer QMLP achieves $0.39$ relative accuracy under PGD ($\varepsilon{=}0.01$), while the 50-layer model drops to $0.13$. This follows directly from the baseline: the 50-layer model already degrades to $32\%$ clean accuracy (vs.\ $54\%$ at 10 layers, Section~\ref{sec:QMLPBaseline}), so gradient-based attacks require minimal perturbation to cause misclassification. MNIST and FGSM follow the same pattern.

Amplitude encoded models collapse to near-zero relative accuracy at $\varepsilon{=}0.10$ across all depths and datasets. Single-shot embedding maps the entire input into a $2^n$-dimensional state before the variational layers, providing a fixed adversarial entry point that deeper layers cannot correct. Small perturbations thus induce large Hilbert-space displacements, making decision boundaries trivially crossable.

At $\varepsilon{=}0.01$, CMLP clearly dominates, retaining above $0.96$ relative accuracy on both datasets, while angle- and amplitude-encoded QMLPs drop to
approximately $0.49$ and $0.30$ respectively at layer~2 on MNIST under FGSM ($0.48$ and $0.27$ under PGD). However, this advantage does not hold uniformly across perturbation strengths on AZ-Class. At $\varepsilon{=}0.10$, angle layer~2 outperforms CMLP under both FGSM and PGD. At $\varepsilon{=}0.15$, the reversal widens: under FGSM, $0.28$ vs.\ $0.13$ at layer~2 and $0.15$ vs.\ $0.13$ at layer~50; under PGD, $0.26$ vs.\ $0.02$ at layer~2, though layer~50 no longer outperforms CMLP ($0.02$ vs.\ $0.02$), indicating that the depth boundary of the reversal is attack-dependent. No reversal is observed on MNIST at any $\varepsilon$, where CMLP retains its advantage throughout. The reversal on AZ-Class reflects the localized qubit-wise structure of angle encoding, which limits cross-feature interference and provides partial resistance at higher perturbations, whereas CMLP's dense representations become more globally exploitable as $\varepsilon$ grows.

In Summary, these results expose a fundamental accuracy-robustness trade-off: amplitude encoding with deep circuits achieves the highest clean accuracy ($93\%$ on MNIST) but collapses under modest perturbations, while shallow angle-encoded models offer a more balanced profile.

\paragraphB{Statistical Analysis}
95\% confidence intervals are computed via the $t$-distribution. At $\varepsilon{=}0.01$, CMLP significantly outperforms
both QMLP variants on all datasets and depths (Welch's $t$-test,
$p{\leq}0.0001$). At $\varepsilon{=}0.10$, CMLP retains a
significant advantage over amplitude encoding ($p{\leq}0.0001$) and
over angle encoding on MNIST ($p{=}0.0001$), but angle layer~2
significantly outperforms CMLP on AZ-Class under both FGSM
($p{=}0.0006$) and PGD ($p{=}0.0005$). At $\varepsilon{=}0.15$,
this reversal persists and widens at layer~2 ($p{\leq}0.0003$),
and extends to layer~50 under FGSM ($p{=}0.0002$) but not under
PGD ($p{=}0.2609$), confirming the attack-dependent depth boundary
of the reversal. Amplitude encoding collapses to near-zero across
all depths at $\varepsilon{\geq}0.10$ ($p{\leq}0.0001$). Depth
significantly degrades angle encoding robustness on AZ-Class across
all $\varepsilon$ values (L2 vs.\ L50, $p{\leq}0.0003$). The sole
non-significant depth comparison is MNIST angle PGD at
$\varepsilon{=}0.10$ ($p{=}0.2402$), where both depths already
approach floor-level accuracy.

\section{Secure and Robust QML Pipeline Design}
\label{sec:QML_pipeline}

\begin{figure}[!t]
    \centering
    \includegraphics[width=\linewidth]{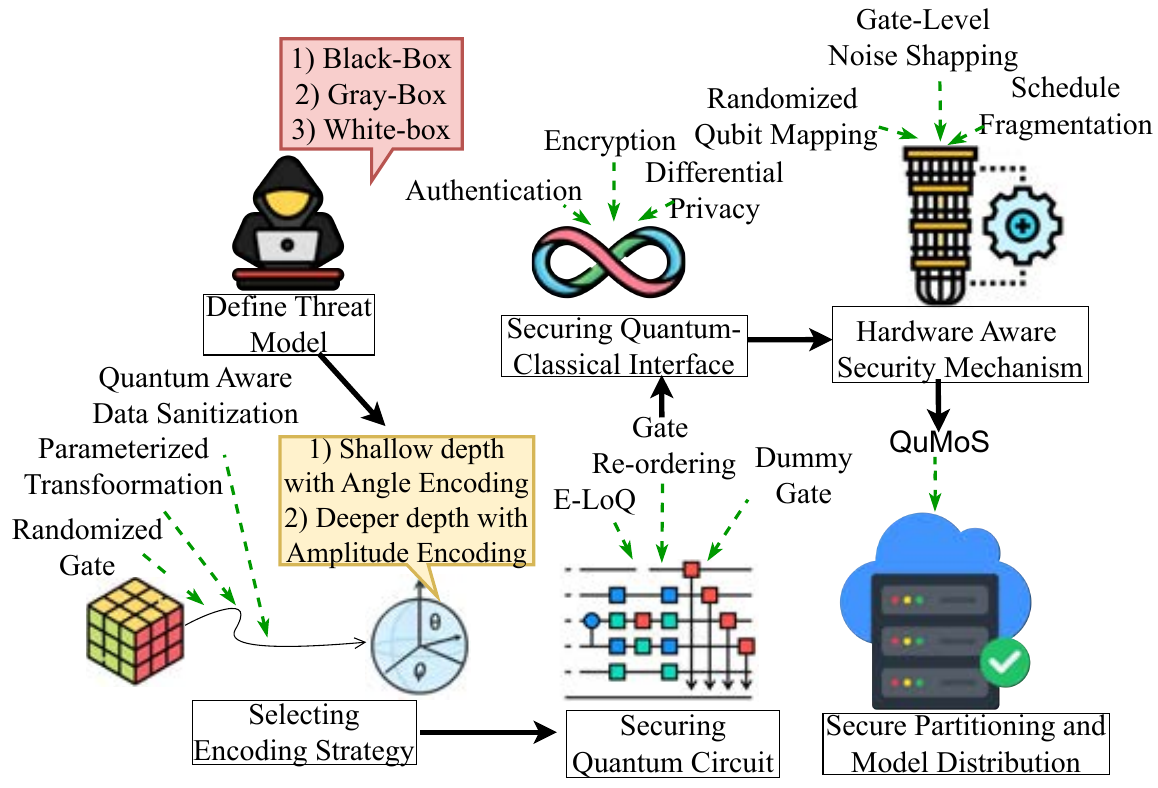}
    \caption{Proposed Secure and Robust QML Pipeline.}
    \label{fig:QML_guideline}
    \vspace{-0.3cm}
\end{figure}

In this section, we present a modular security framework for QML systems grounded in our threat model and the empirical findings of Section~\ref{sec:5-evaluation}. Our results highlight three key lessons that guide the recommendations below: classical defenses do not transfer directly to QML, noise is an unreliable and asymmetric passive defense, and encoding choice is the most consequential architectural security decision. Figure~\ref{fig:QML_guideline} maps the main defenses to each stage of the QML pipeline.


\paragraphB{Defining the Adversarial Threat Model}
Securing a QML pipeline starts with a clear threat model that defines the adversary’s access, capabilities, and goals. In QML, this means identifying which components such as encoded data, transpiled circuits, or pulse-level signals—can be observed or modified, and linking them to realistic attack vectors. A threat model aligned with our taxonomy (Section~\ref{sec:taxonomy}) therefore provides the basis for secure design, enabling realistic robustness assessment and avoiding both over- and under-engineered defenses.

\paragraphB{Encoder-Level Security Considerations}
The encoding stage is the most security-critical interface in our evaluation. QUID~\cite{Kundu-poisoning-2025} exploits encoded-state structure to poison training labels with minimal overhead, with amplitude encoding showing near-complete attack success due to strong class separation (Section~\ref{sec:quid-eval}). Under FGSM and PGD, amplitude-encoded models also collapse at $\epsilon=0.10$ across all depths, since single-shot embedding gives perturbations a fixed entry point (Section~\ref{sec:PGD-FGSM}). By contrast, shallow angle encoding offers a better accuracy-robustness tradeoff through localized qubit-wise mapping, making encoding choice a key security decision in QML. To mitigate these risks, encoding should include quantum-aware validation~\cite{jiang2020towards,Yu2022StateVerification,Govindankutty2025Superposition}, randomized or obfuscated schemes~\cite{gong2024enhancing,Das2023RandomizedRG}, and classical-quantum consistency checks~\cite{kit_2025}.

\paragraphB{Securing Quantum Circuit Architecture}
VQCs are vulnerable to extraction and reverse engineering~\cite{Ghosh2024Imitation}. Our QTrojan~\cite{chu2023qtrojan} evaluation showed that an attacker with full circuit access can insert backdoor gates that remain inactive until triggered, redirecting predictions without access to training data. Although QTrojan (Section~\ref{sec:qtrojan-eval}) was ineffective in our multi-class setting, it still shows that circuit-level access creates a fundamental white-box risk that data-level or noise-based defenses cannot address. To protect circuit confidentiality, quantum logic locking (QLL)~\cite{Topaloglu2023logic-locking} and E-LoQ~\cite{liu2025eloqenhancedlockingquantum} enforce key-dependent circuit behavior, while obfuscation through gate reordering and dummy-gate insertion~\cite{Das2023RandomizedRG,raj2025quantum,Suresh2022Obfuscation} makes static analysis and reverse engineering harder.

\paragraphB{Hardware-Aware Security Mechanisms}
Our results show that noise is an unreliable passive defense: it weakens QUID~\cite{Kundu-poisoning-2025} on amplitude encoding by disrupting encoded-state structure (Section~\ref{sec:quid-eval}), but does not stop the Huang backdoor~\cite{huang2023backdoor}, whose trigger remains embedded in the model weights (Section~\ref{sec:huang-eval}). This asymmetry calls for active hardware-level defenses rather than reliance on incidental noise. Useful measures include randomized qubit mapping~\cite{choudhury25crosstalk}, instruction reordering and staggered scheduling~\cite{Xie2022Crosstalk,Harper2025Crosstalk}, dynamical decoupling~\cite{Mehra2024Crosstalk,choudhury25crosstalk}, pulse-channel verification and noise shaping~\cite{xu2025security}, and hardware noise fingerprinting for tamper detection~\cite{Choudhury2024SideChannel}.

\paragraphB{Secure Partitioning and Model Distribution} 
To mitigate centralized deployment risks, QML models should adopt partitioning strategies like QuMoS~\cite{wang2023qumos}, which divide functionality across multiple quantum backends via isolated sub-circuits that communicate through secure classical channels. This architectural isolation is particularly important given our label-flipping results (Section~\ref{sec:label-flipping}), which showed that classical defenses such as label smoothing~\cite{Szegedy2015RethinkingTI} provide no meaningful protection for QMLP- underscoring that reducing the attack surface at deployment is necessary when algorithmic defenses alone are insufficient.

\section{Practical Challenges and Limitations}

Due to the substantial computational demands of QML models, we were unable to evaluate several configurations in this study. First, we did not assess \emph{randomized encoding}~\cite{gong2024enhancing} as a defense mechanism. Preliminary tests showed that randomized gates significantly increased circuit depth and simulation overhead, making this approach impractical even with high-memory GPUs. We also excluded all attacks on \emph{50-layer angle-encoded} QMLP models, except the white-box evasion attacks. These deep circuits require repeated data re-uploading across variational layers, resulting in prohibitive runtime and memory usage under our hardware constraints. The white-box evasion attacks (FGSM and PGD) were not evaluated under noisy settings, as these experiments were conducted on the full datasets, making adversarial perturbation computation under noise computationally prohibitive. Additionally, results for 5-layer and 10-layer configurations could not be included in the paper due to space constraints; however, they are available in our GitHub repository.

Finally, we restricted our experiments to 10-class MNIST classification and 23-class AZ-Class classification. Scaling to higher class counts caused exponential increases training time, exceeding the practical time budget within our available computational resources.

\section{Conclusion}

In this work, we present the first comprehensive systematization of adversarial threats in quantum machine learning (QML), encompassing both classical-inspired and quantum-native attack vectors. Our results show that robustness of QML systems is significantly influenced by encoding strategies, circuit depth, and noise characteristics. Existing defenses remain limited in scope and are often constrained to simulation environments. As QML systems move closer to real-world deployment, there is a critical need to develop quantum-native robustness techniques, hardware-aware circuit designs, and formal threat models tailored to quantum architectures. 

\section{Ethics considerations}
None

\bibliographystyle{IEEEtran}
\bibliography{main-references}

\appendices

\clearpage 

\section{Meta-Review}

The following meta-review was prepared by the program committee for the 2026 IEEE Symposium on Security and Privacy (S\&P) as part of the review process as detailed in the call for papers.

\subsection{Summary}
This paper provides an overview of attacks against machine learning algorithms specifically designed for quantum computation. Representative samples from different categories of attacks are implemented and tested.

\subsection{Scientific Contributions}

1. Independent Confirmation of Important Results with Limited Prior Research.\\

6. Provides a Valuable Step Forward in an Established Field.

\subsection{Reasons for Acceptance}
\begin{enumerate}
    \item This paper provides independent confirmation of important results with limited prior research. It pulls together a review of attacks against quantum machine learning algorithms, implementing samples from each category of attack to provide additional empirical insights.
    \item This paper provides a valuable step forward in an established field. While attacks on machine learning are not new, this paper provides a systematization of those attacks as applied in quantum computing, comparing with their counterparts in classical machine learning where appropriate. Empirical evaluation is used to confirm and supplement the results from prior work.
\end{enumerate}

\end{document}